\documentclass[letterpaper,11pt]{article}
\usepackage[utf8]{inputenc}
\usepackage{graphicx}
\usepackage[margin=0.75in]{geometry}
\usepackage{cite}
\usepackage{amsmath}
\usepackage{jheppub}
\usepackage{xspace}
\usepackage{booktabs} % Essential for professional spacing and \cmidrule
\usepackage{mathrsfs}
\usepackage{float}
\usepackage{multirow} 

\author[a]{Adam Berger}
\author[a]{Riccardo Catena}
\author[b]{Jan Conrad}
\author[a]{Taylor R. Gray}
\affiliation[a]{Chalmers University of Technology, Department of Physics and Astronomy, SE-41296, Gothenburg, Sweden}
\affiliation[b]{The Oskar Klein Centre, Department of Physics,
Stockholm University, AlbaNova, SE-10691, Stockholm, Sweden}

\emailAdd{aberger@student.chalmers.se}
\emailAdd{catena@chalmers.se}
\emailAdd{conrad@fysik.su.se}
\emailAdd{taylor.gray@chalmers.se}

\newcommand{\micromegas}{\texttt{micrOMEGAs}\xspace}
\newcommand{\madgraph}{\texttt{MadGraph}\xspace}

\newcommand{\Ap}{A^{\prime}}

\newcommand{\figref}[1]{Fig.~\ref{#1}}
\newcommand{\secref}[1]{Section~\ref{#1}}
\newcommand{\appref}[1]{Appendix~\ref{#1}}
\newcommand{\eqnref}[1]{Eq.~\ref{#1}}
\newcommand{\tabref}[1]{Table~\ref{#1}}

\abstract{Light dark matter (DM) is a compelling scenario for the observed relic abundance, with accelerator-based searches as a powerful discovery strategy. The upcoming Light Dark Matter eXperiment (LDMX) is designed to probe light DM by measuring the energy and transverse momentum of recoil electrons in high-intensity electron–nucleus collisions.
We evaluate the discovery potential of LDMX to light DM at four benchmark points along the thermal relic target for complex scalar DM mediated by a dark photon, for different background assumptions, and assess its model selection power at a representative benchmark.
We find that LDMX has strong projected 5$\sigma$ discovery potential along the relic target across both background scenarios for certain benchmarks, and we further compute projected 90\% C.L. exclusion limits assuming no measured signal events.
The normalization and shape of the two-dimensional recoil electron distribution encodes the couple strength and dark photon mass, respectively, enabling parameter inference in the event of a signal excess. We perform parameter estimation on simulated data and find that both parameters are recovered within their uncertainties along the relic target across different background scenarios.
Next, we assess whether the data can distinguish between competing dark sector hypotheses, in particular, dark photons with additional higher electromagnetic moment interactions.
We demonstrate that model comparison using the Bayes factor allows dark photon model hypotheses to be statistically distinguished, with the two-dimensional analysis affording substantially greater discriminating power than an energy-only analysis.
These results are obtained within a likelihood-based statistical framework for LDMX, incorporating signal and background modelling with their associated systematic uncertainties and employing both frequentist and Bayesian methods.
The framework is designed for direct application to real LDMX data.}

\title{Light Dark Matter Discovery Potential and Model Selection at LDMX}

\begin{document}

\maketitle

\section{Introduction}

Accelerator-based experiments provide competitive sensitivity to light, sub-GeV dark matter (DM)~\cite{Caputo:2026pdw,Balan:2024cmq}, complementary to direct detection searches via electron and nuclear recoils~\cite{DAMIC-M:2025luv,PandaX:2025rrz,DarkSide:2022knj,CRESST:2024cpr,Krnjaic:2025noj}. A well-motivated theoretical framework for light DM, invokes a dark sector coupled to the Standard Model (SM) through a massive vector mediator, the dark photon, which kinetically mixes with the ordinary photon with strength $\epsilon$~\cite{Holdom:1985ag}. In fixed-target and beam-dump experiments, the dark photon can be produced via dark bremsstrahlung or meson decay and subsequently decay invisibly to a DM pair, leaving a distinctive missing-energy or missing momentum signature.

The Light Dark Matter eXperiment (LDMX) is a proposed electron fixed-target missing momentum experiment, operating with an 8 GeV electron beam \cite{LDMX:2025bog}. The experiment is designed to search for light DM produced via dark bremsstrahlung, in which a beam electron scatters off a thin tungsten target and radiates a dark photon that subsequently decays invisibly to a DM pair. The resulting signal signature is a single recoil electron carrying anomalously low energy and large transverse momentum, accompanied by a complete absence of activity in the downstream calorimeters. A silicon tracker upstream and downstream of the target reconstructs the recoil electron kinematics, while electromagnetic and hadronic calorimeters provide the veto on SM processes — in particular electro-nuclear and photo-nuclear interactions. With a projected $4\times10^{14}$ electrons on target (EOT) in an initial phase (Phase I) and up to $10^{16}$ in a subsequent phase (Phase II), LDMX is designed to improve on existing constraints by orders of magnitude in coupling, providing broad sensitivity to the thermal freeze-out targets for a wide range of mediator and DM masses in the MeV–GeV range~\cite{Berlin:2018bsc}.

If LDMX observes an excess of events inconsistent with the background-only hypothesis, it would represent a landmark discovery of a new massive dark sector state. Beyond a mere detection, the shape of the two-dimensional recoil distribution carries information about the mediator mass and kinetic mixing strength, opening the prospect of characterizing the new particle's properties directly from the data. Extracting these results in a statistically rigorous way, and understanding the conditions under which discovery and parameter inference are feasible, requires a full likelihood framework.

We present a likelihood-based statistical framework for LDMX built partly on the statistical methodology of~\cite{Likelihoods}, exploiting the two-dimensional distributions of recoil electron energy and transverse momentum, and incorporating both signal and background modelling with their associated systematic uncertainties. The framework is applied to simulated data across two representative background scenarios: one with $N_\text{BG} = 1$ background event, and the other with $N_\text{BG} = 100$ background events. These background scenarios bracket approximately the range of conditions expected in LDMX running, and show the impact of different background levels. We run full toy pseudo-experiments to compute the test statistic distributions since the low bin counts render asymptotic approximations unreliable, and compute median quantities such as the exclusion $p$-value and the discovery significance as a projection of the sensitivity. Accurate modelling of both signal and background distributions, and a thorough understanding of background normalization uncertainties, are shown to be critical ingredients, underscoring their importance in the design of future LDMX analyses. 

We assess the projected sensitivity and predictive power of LDMX, assuming Phase II luminosity of $10^{16}$ EOT, by employing frequentist profile likelihood methods for setting exclusion limits, claiming discovery, and performing parameter estimation on the kinetic mixing strength $\epsilon$ and mediator mass $m_{A'}$, while Bayesian inference is used for posterior-based uncertainty quantification, parameter estimation, and model discrimination
via the Bayes factor. See also~\cite{DMsignals_MM} for a complementary kinematic approach to signal characterization at such experiments.

We situate our results against four benchmark points on the thermal relic targets that remain viable after the most recent direct detection constraints~\cite{Krnjaic:2025noj}. We consider a scalar DM candidate across mediator to DM mass ratios $\mathcal{R} \equiv m_{A'}/m_\text{DM} = 2.5$ and $ \mathcal{R} = 2.2$, which remain unconstrained by current bounds. These benchmarks sit just above the threshold where the dark photon goes on-shell and the DM annihilation cross section is resonantly enhanced near freeze-out, making them the most challenging regime for exclusion and thus the most relevant surviving benchmarks.

Although the framework is demonstrated here on simulated data, it is designed to be straightforwardly adapted to real LDMX data once available, and can naturally be extended to a combined likelihood incorporating results from other light DM probes, enabling joint constraints across the broader experimental program. Furthermore, we employ the dark photon scenario in which it decays invisibly on-shell to DM states, however this framework can apply generically to any dark sector scenario beyond kinetic mixing where couplings to electrons are present such as \cite{Banerjee:2025ejz}. 

The dark sector theory is outlined in \secref{section:theory}, which introduces the dark photon mediator, the scalar DM candidate used for defining benchmarks, and the modelling of dark bremsstrahlung for the expected signal distributions. \secref{section:Likelihood} describes the likelihood-based statistical formalism used throughout this study for both frequentist and Bayesian analyses. The frequentist results -- including the projected exclusion bounds, discovery significance, and maximum likelihood parameter estimation -- are detailed in \secref{section:frequentist}. The Bayesian analysis follows in \secref{section:bayesian}, covering posterior distributions, uncertainties on the inferred model parameters, in addition to model comparison for different dark photon models. We conclude in \secref{section:conclusion}.

\section{Dark Sector Theory}
\label{section:theory}

In this section we summarize the dark photon and the benchmark scenario of complex scalar DM. Next we introduce the dark higher order electromagnetic moment dark photon models considered in the context of signal discrimination. Finally, we discuss the signal and background modelling.

\subsection{The Dark Photon}
\begin{figure}
    \centering
\includegraphics[width=0.7\linewidth]{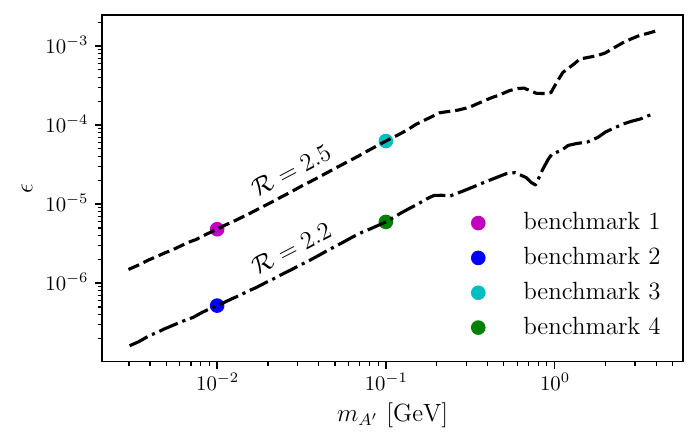}
    \caption{Thermal relic targets, on the kinetic mixing $\epsilon$ vs dark photon mass plane, for complex scalar DM with $\mathcal{R}\equiv \frac{m_{\Ap}}{m_\text{DM}} =$ 2.5 (dashed) and 2.2 (dash dotted), where the observed relic density is obtained assuming freeze-out \cite{Berlin:2018bsc}. Four benchmarks used throughout this work are plotted. Analogous thermal targets are computed, for the same $m_{A'}$ and $\mathcal{R}$, for the DEM models discussed in \secref{subsubsection:DEM}.}
    \label{fig:BMs}
\end{figure}

Missing momentum/energy searches such as LDMX are sensitive to invisible massive states that are produced through electron beam interactions with the target. In this work, we consider the production of the dark photon, $A'$, the gauge boson of a new $U(1)'$ gauge group, which acts as a mediator between the DM and the SM \cite{Holdom:1985ag,Fabbrichesi:2020wbt,Caputo:2026pdw}. The dark photon obtains its mass through a dark Higgs or the Stueckelberg mechanism \cite{Stueckelberg:1938hvi}. We consider dark photon masses in the 1 MeV to 1 GeV range \cite{Boehm:2003hm} due to experimental relevance.
Below 1 MeV, dark sector particles are subject to strong constraints from cosmology if the particle is thermal \cite{Ibe:2019gpv,Sabti:2019mhn}. Furthermore, LDMX loses sensitivity below $m_{A'} \sim$ 1 MeV since the dark photon is produced nearly collinear with the beam, resulting in recoil electrons with insufficient transverse momentum and missing energy to pass the signal selection and trigger requirements. Therefore, we restrict ourselves to the dark photon mass range MeV - GeV.
The upper mass bound is set by the available center-of-mass energy in electron-nucleus collisions at the LDMX beam energy, above which direct detection experiments searching for nuclear recoils are more suitable. The coupling to SM fermions come from the kinetic mixing with the SM hyper-charge \cite{Holdom:1985ag},
\begin{equation}
    \mathscr{L}_\text{KM} = -\frac{\epsilon}{2 \cos \theta_W}F'_{\mu\nu}B^{\mu \nu},
\end{equation}
where $\epsilon$ is the kinetic mixing parameter, $\theta_W$ is the weak mixing angle, $F'_{\mu \nu}$ is the field strength tensor of the $U(1)'$ gauge theory, and $B^{\mu \nu}$ is the field strength tensor of the SM hypercharge (both before field redefinitions). This kinetic mixing operator can be generated by a vacuum polarization like diagram with new heavy fields carrying both SM and dark charges \cite{Rizzo:2018vlb}. After field redefinitions and transformation to the mass eigenstates we have the Lagrangian,
\begin{equation}
    \mathscr{L}_{A'} = -\frac{1}{2} m_{\Ap}^2 {\Ap}^{\mu} \Ap_\mu - \frac{1}{4}{\Ap}^{\mu \nu} \Ap_{\mu \nu} - \epsilon e A'^\mu \sum_f q_f \bar{f} \gamma_\mu f,
    \label{eq:L_A'}
\end{equation}
where $A'^\mu$ is the dark photon vector field, $A'^{\mu\nu}$ is its field strength tensor, $f$ are SM fermions with corresponding charge $q_f$, and $e$ is the elementary charge. For the sake of computing meaningful benchmarks on the parameters $m_{A'}$ and $\epsilon$, we consider a complex scalar DM candidate $\chi$ with the following Lagrangian\footnote{See \cite{Berlin:2018bsc,Banerjee:2025ejz,Catena:2023use} for other DM theory scenarios probed by LDMX and other accelerator experiments.},
\begin{equation}
    \mathscr{L}_\text{DM} = |\partial_\mu \chi|^2 - m_{\text{DM}}^2|\chi|^2 + ig_{\text{D}}A'^\mu \big[ \chi^* \left(\partial_\mu \chi \right) - \left(\partial_\mu \chi^* \right) \chi \big] - g_D^2 A'^\mu A'_\mu |\chi|^2,
\end{equation}
where the dark fine-structure constant commonly used in the literature is $\alpha_D \equiv \frac{g_D^2}{4\pi}$, which we take to be $\alpha_D=0.5$ throughout this work. The ratio of the masses, $\mathcal{R} \equiv \frac{m_{\Ap}}{m_\text{DM}}$, has been commonly fixed at $\mathcal{R} = 3$ in previous studies. Recent null-results from electron recoil direct detection experiments DAMIC-M \cite{DAMIC-M:2025luv} and PANDAX-4T \cite{PandaX:2025rrz}, in addition to previous results from DarkSide-50 \cite{DarkSide:2022knj} and CRESST-III \cite{CRESST:2024cpr}, now place strong constraints on sub-GeV DM -- excluding complex scalar DM  with $\mathcal{R} \gtrsim 3$ 
\cite{Krnjaic:2025noj}. We thus take as benchmarks the $\epsilon$ values on the $\mathcal{R}=2.5$ and $\mathcal{R}=2.2$ thermal targets\footnote{The parameters which lie on the \textit{thermal target} are those that predict the observed relic abundance by Planck \cite{Planck:2018vyg} via the freeze-out mechanism \cite{Gondolo:1990dk}.} from \cite{Berlin:2018bsc} at $m_{A'} = \{ 0.01,0.1 \}$ GeV, illustrated in \figref{fig:BMs}

\subsubsection{Dark Higher Order Electromagnetic Moments}
\label{subsubsection:DEM}

In addition to the kinetic mixing interaction between the dark photon and SM fermions, other types of interactions of higher dimensionality can be generated from loop diagrams of heavy \textit{portal matter} fields and gauge bosons of new symmetry groups\footnote{The exact UV physics that generates these effective interactions is beyond the scope of this study, however work in this direction has been carried out in \cite{Rizzo:2018vlb,Rizzo:2023djp,Rizzo:2024bhn}.} \cite{Rizzo:2018vlb}. Interestingly, these interactions predict different kinematic distributions of the recoil electron due to the momentum dependence within the Lorentz structures \cite{Catena:2025fsl}. The set of these dark electromagnetic moment (DEM) interactions are,
\begin{equation}
\begin{aligned}
    \mathscr{L}_{\text{DEM}} = & \sum_f \Bigg(  
    - \mu_f \partial_\nu \left( \bar{f} \sigma^{\mu \nu} f \right) A'_\mu -i d_f \partial_\nu \left( \bar{f} \sigma^{\mu \nu} \gamma^5 f \right) A'_\mu \\ 
    &- b_f  \left[ \partial^\nu \partial_\nu \left(\bar{f}\gamma^\mu f \right) - \partial^\mu \partial_\nu \left(\bar{f}\gamma^\nu f \right) \right] A'_\mu \\ 
    &+ a_f \left[ \partial^\nu \partial_\nu \left(\bar{f}\gamma^\mu \gamma^5 f \right) - \partial^\mu \partial_\nu \left(\bar{f}\gamma^\nu \gamma^5 f \right) \right] A'_\mu \Bigg),
\label{eq:lagrangian}
\end{aligned}
\end{equation}
where the first term is the magnetic dipole term with coupling $\mu_f$ denoted M, the second the electric dipole with $d_f$, E, the third the charge radius with $b_f$, C, and the last the anapole moment with $a_f$, A. We define the parameter $g_f$ representing the coupling strength for each model,
\begin{equation}
g_f = 
\begin{cases}
    \epsilon & \text{Kinetic Mixing} \\
    \mu_f/e & \text{Magnetic Dipole} \\
    d_f/e & \text{Electric Dipole} \\
    b_f/e & \text{Charge Radius} \\
    a_f/e & \text{Anapole Moment}.
\end{cases}
\label{eq:g_f}
\end{equation}
Notice that the units of $g_f$ vary depending on the model \cite{Catena:2025fsl}.

To quantify the discriminating power of LDMX between different signal hypotheses, we consider each interaction separately: kinetic mixing (KM) (the third term in \eqnref{eq:L_A'}), M, E, C, and A. We compute the corresponding thermal target benchmarks of each dark photon model at $\mathcal{R} = 2.5$ and $\mathcal{R} = 2.2$ at $m_{A'}=\{0.01,0.1\}$ GeV, using \micromegas \cite{Belanger:2013oya} computed thermally averaged cross sections and our in-house Boltzmann solver based off of \cite{Gondolo:1990dk}.

\subsection{Signal and Background Modelling}
\label{section:Theory_SB}
We present the methodology used to compute the expected signal and background signatures in the two-dimensional space of recoil electron energy $E_{e^-_f}$ and transverse momentum  $|p_T|_{e^-_f}$, which enter into the likelihood framework presented in \secref{section:Likelihood}. We consider the LDMX Phase II specifications \cite{LDMX:2025bog}, with an 8 GeV electron beam and $10^{16}$ electrons on target ($N_e$). Both signal and background distributions are smeared due to the detector resolution, as described in \appref{section:smear}. The simulated events are binned in two-dimensional bins spanning $|p_T|_{e^-_f} \in [0,2]$ GeV and $E_{e^-_f} \in [0.05,8]$ GeV. The lower energy threshold of $E_{e^-_f} = 50$ MeV is imposed to account for the degradation of tracking acceptance in this region~\cite{LDMX:2018cma}, and we additionally require the recoil electron to be within $\pm 40^\circ$ of the beam axis \cite{LDMX:2025bog}. or equivalently $ \sin^{-1}(|p_T|_{e^-_f}/E_{e^-_f}) < 40^\circ$. 
The upper limits on both $|p_T|_{e^-_f}$ and $E_{e^-_f}$ are motivated by the beam energy and the negligible expected yield of events with $|p_T|_{e^-_f} > 2$ GeV.
The LDMX collaboration reports the energy cut $E_{e^-_f}<3.16$ GeV from the trigger requirement \cite{LDMX:2025ixw,LDMX:2025bog}, therefore we investigate the impact of this cut on the exclusion and discovery reach. Within our analysis, this energy cut reduces the overall background rate by approximately a factor of 10, and the signal rate by only a factor of $\sim 1.6$ for small $m_{A'}$, and $\sim 1.1$ for large $m_{A'}$.
Throughout this work we use a grid of $50\ |p_T|_{e^-_f} \times 50\ E_{e^-_f}$ bins, yielding 2500 bins in total. This binning is fine enough to resolve shape differences between signal hypotheses across the range of dark photon masses and coupling models considered, while keeping the computational cost of the analysis manageable.

\subsubsection*{Signal}
\begin{figure}
    \centering
\includegraphics[width=0.8\linewidth]{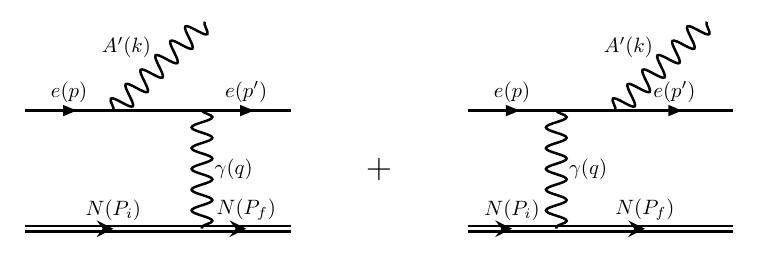}
    \caption{Dark photon bremsstrahlung, where an electron scatters off a target nucleus $N$ and radiates a dark photon.}
    \label{fig:feynman}
\end{figure}
Dark photons can be produced at LDMX through beam electron target collisions through an analogous process to ordinary bremsstrahlung, dark bremsstrahlung \cite{Liu:2017htz}, drawn in \figref{fig:feynman}, followed by the invisible decay of the dark photon to DM,
\begin{equation}
    e^- N \to e^- N \Ap, \: \Ap \to \chi \chi^*.
\end{equation}
The corresponding signature is the missing energy, and the non-zero momentum component perpendicular to the beam, of the recoil electron, with an absence of any other activity in the calorimeters. 

\begin{figure}
    \centering
\includegraphics[width=0.9\linewidth]{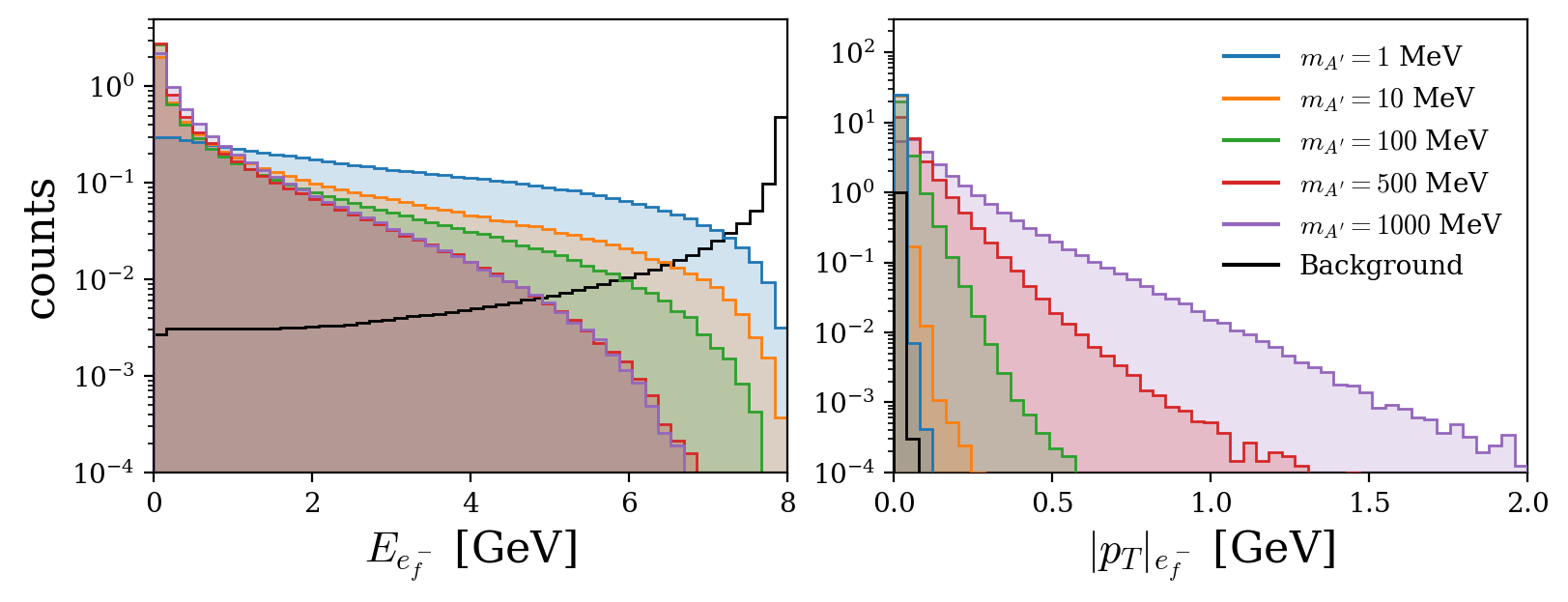}
    \caption{Normalized distributions of the recoil electron's energy (left) and transverse momentum (right) in dark photon bremsstrahlung, for varying dark photon masses, $m_{\Ap}$. The background is drawn in black.}
    \label{fig:histograms}
\end{figure}

\begin{figure}
    \centering
\includegraphics[width=0.9\linewidth]{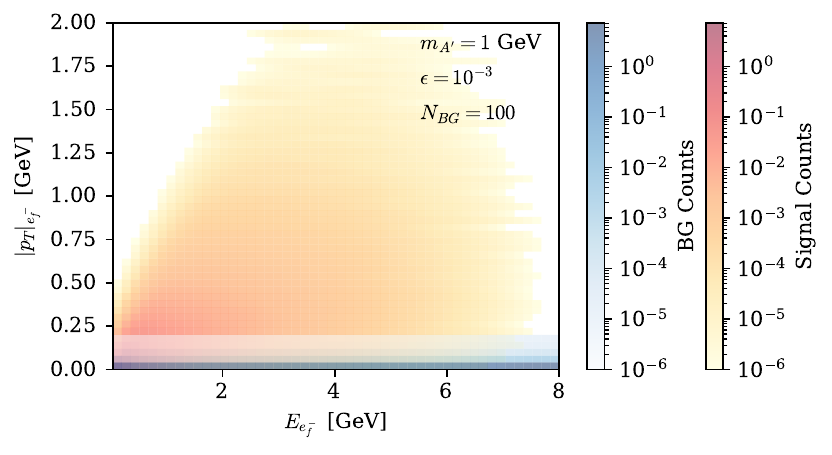}
    \caption{Two-dimensional distributions of the recoil electron energy $E_{e_f^-}$ and transverse momentum $|p_T|_{e_f^-}$ for the background (blue) and for the signal at a representative dark photon mass and kinetic mixing (yellow/orange). The binning used in the likelihood construction is shown, with 50$\times$50 bins.}
    \label{fig:2Dhistograms}
\end{figure}

We simulate dark photon production using \texttt{MadGraph5\_aMC\@NLO} \cite{Alwall:2014hca} for varying $\Ap$ masses, yielding the expected binned signal distributions in $E_{e^-_f}$ and $|p_T|_{e^-_f}$ presented in \figref{fig:histograms}. The associated two-dimensional distributions used directly in the likelihood framework are displayed in yellow/red in \figref{fig:2Dhistograms}.

The total number of expected dark photon events is,
\begin{equation}
    N_{\Ap}(g_f, m_{\Ap}) =  \sigma_{\Ap}(g_f, m_{\Ap}) \frac{T X_0 N_e N_0}{A} ,
\end{equation}
where $\sigma_{\Ap}$ is the cross section for the process in \figref{fig:feynman} which is obtained using \madgraph, $T$ is the number of radiation lengths taken to be 0.1, $X_0$ is the radiation length of the target which is  6.76 $\text{g}/\text{cm}^2$ for tungsten, $A$ is the atomic mass, and $N_0$ is Avogadro's number \cite{Izaguirre_2015}. The signal distributions are thus normalized to $N_{A'}$ for a given $m_{A'}$ and $g_f$.

\subsubsection*{Background}
The main source of background for LDMX are bremsstrahlung photons which go undetected, for example due to photo-nuclear interactions producing neutral hadrons \cite{LDMX:2023zbn}. We take the photo-nuclear $|p_T|_{e^-_f}$ distribution of the recoil electron from Fig. 6 of \cite{LDMX:2023zbn}, and the inclusive single electron background $E_{e^-_f}$ distribution from Fig. 10 of \cite{LDMX:2018cma} as the background for this study. The two-dimensional background distribution is then constructed as the outer product of these two one-dimensional distributions, assuming that $|p_T|_{e^-_f}$ and $E_{e^-_f}$ are uncorrelated for background events. This assumption serves as an approximation for the purposes of this study.\footnote{A more dedicated characterization of the two-dimensional background distribution would be valuable for a full analysis. While significant work has already been carried out in this direction within the LDMX collaboration \cite{LDMX:2023zbn, LDMX:2018cma}, a complete understanding of the overall normalization at $10^{16}$ EOT and the correlated $(|p_T|_{e^-_f}, E_{e^-_f})$ background shape — including potential contributions from additional processes — remains an interesting avenue for future study.} We take as benchmarks $N_{\text{BG}} = 1$ and $100$ for the overall normalization of the background, the explore the effect of varying background levels. The background is plotted in blue in \figref{fig:2Dhistograms}.

\section{Likelihood Formalism}
\label{section:Likelihood}

We present the binned likelihood in the two-dimensional space of recoil electron energy $E_{e_f^-}$ and transverse momentum $|p_T|_{e_f^-}$, incorporating both signal and background uncertainties through nuisance parameters.

Given the predicted number of events in bin $i$, $\nu_i = S_i + B_i$ (the predicted number of signal events $S_i$ plus background events $B_i$), the probability of measuring $d_i$ events in this bin is described by the Poisson distribution,
\begin{equation}
    f_P(d_i|\nu_i) = \frac{\nu_i^{d_i}}{d_i!}e^{-\nu_i}.
\end{equation}
The number of signal events for bin $i$, $S_i$, is a function of the DM model parameters $\epsilon$, where $S_i \propto \epsilon^2$, and $m_{\Ap}$, with a non-trivial dependence. See \secref{section:Theory_SB} for the details.
Therefore, we can write the likelihood of the data given the model parameters ($\epsilon,m_{\Ap}$) as a product of the Poisson probabilities of each bin, assuming that the observed bin counts are independent.

Since each event is characterised by both $E_{e^-_f}$ and $|p_T|_{e^-_f}$, these two observables can be combined into a joint two-dimensional likelihood, where the product runs over bins in the $(E_{e^-_f},|p_T|_{e^-_f})$ plane as shown in \figref{fig:2Dhistograms}. This two-dimensional approach retains more discriminating information than treating the two variables separately.
The signal (in yellow/orange) and the background (in blue) are computed and binned as described in \secref{section:Theory_SB}.

Two standard approaches exist for treating systematic uncertainties: in a frequentist framework, nuisance parameters are introduced into the likelihood and subsequently profiled (maximised over) \cite{Lista:2016chp}, while in a Bayesian framework they are assigned prior distributions and marginalised over. See \cite{Conrad:2002kn} for a combination of the approaches in a frequentist analysis. We adopt both approaches: the constraint terms $\mathcal{P}(\theta | \sigma^2)$ serve as penalty terms in the profile likelihood, and as priors in the Bayesian posterior.
We introduce the nuisance parameters $\theta_S$ and $\theta_B$ into the likelihood, where $\theta_S$ and $\theta_B$ are multiplicative scale factors with nominal value 1, such that at 1 the baseline signal/background predictions are recovered. The constraint terms are chosen to be log-normal distributions, since they are positive-definite.
The likelihood is then,
\begin{equation}
\begin{aligned}
    \mathcal{L}(\vec{d}|\epsilon,m_{\Ap},\theta_S,\theta_B) &= \prod_{i}^{N_E} \prod_j^{N_{p_T}} \frac{\big[ S_{ij}(\epsilon,m_{\Ap})\theta_S + B_{ij}\theta_B\big]^{d_{ij}}}{d_{ij}!} e^{- \big[ S_{ij}(\epsilon,m_{\Ap})\theta_S + B_{ij}\theta_B\big] } \\ & \times \mathcal{P}(\theta_B| \sigma_B^2) \mathcal{P}(\theta_S | \sigma_S^2),
    \label{eq:likelihood}
\end{aligned}
\end{equation}
where $\mathcal{P}(\theta | \sigma)$ are log-normal distributions defined by,
\begin{equation}
    \mathcal{P}(\theta | \sigma^2) = \frac{1}{\sqrt{2 \pi} \sigma \theta } \exp \bigg[ - \frac{1}{2} \left( \frac{\ln \theta}{\sigma}\right)^2 \bigg].
\end{equation}
We take $\sigma_S$ and $\sigma_B$ to be 0.1, consistent with the $\sim$10\% Monte Carlo integration uncertainty reported by \madgraph, which is treated as representative of the overall signal and background normalization uncertainty.
In the case of the Bayesian analysis, the log-normal distributions are instead implemented as the priors of $\theta_B$ and $\theta_S$.

\subsection{Frequentist Statistical Framework} 

We perform a Frequentist analysis on the kinetic mixing dark photon model, determining the projected exclusion, discovery reach, and maximum likelihood parameter reconstruction power.

\subsubsection{Exclusion}

In this section we describe the methodology used to determine the projected exclusion sensitivity of LDMX: the set of values of $\epsilon$ as a function of $m_{\Ap}$ that can be excluded at a given confidence level (C.L.).

The $p$-value for the exclusion of a signal hypothesis quantifies the probability of observing a test statistic at least as extreme as the one observed, assuming that signal hypothesis is true. A signal hypothesis is considered excluded at the C.L. $1-\alpha$ when $p \leq \alpha$; throughout this work we adopt $\alpha = 0.1$, corresponding to 90\% C.L. exclusion.

Since we are computing the projected exclusion limits --- that is, the sensitivity in the absence of an observed signal --- we replace the observed value of the test statistic with the median of 
its distribution under the background-only hypothesis, $\text{med}[f(q_\epsilon | 0)]$.
The expected exclusion $p$-value at a given $(\epsilon,m_{\Ap})$ is therefore,
\begin{equation}
    p = \int_{\text{med}[f(q_\epsilon|0)]}^{\infty} f(q_\epsilon|\epsilon)\, dq_\epsilon,
\end{equation}
where $f(q_\epsilon|\epsilon)$ is the distribution of the test statistic with data simulated under the signal-plus-background hypothesis with coupling $\epsilon$, obtained from pseudo-experiments. 
The lower limit of integration, $\text{med}[f(q_\epsilon|0)]$, is the median of the test 
statistic distribution under the background-only hypothesis ($\epsilon' = 0$), and serves as a proxy for the expected observation in the absence of a signal.

The test statistic is defined as\footnote{The test statistic and likelihood are commonly written as functions of the signal strength parameter $\mu$, as is written in \cite{Likelihoods}, where the total number of expected events is $\mu S_i + B_i$. We instead use the model parameter $\epsilon$ directly in our notation, since the signal yield is a function of the kinetic mixing parameter $\epsilon$.} \cite{Likelihoods},
\begin{equation}
    q_\epsilon =
    \begin{cases}
    -2 \log \frac{\mathcal{L}\big(\vec{d}(\epsilon')\: |\: \epsilon, \: \hat{\hat{\theta}}_S, \: \hat{\hat{\theta}}_B \: ; \: m_{\Ap} \big)}{\mathcal{L}\big(\vec{d}(\epsilon')\:|\:\hat{\epsilon},\: \hat{\theta}_S,\: \hat{\theta}_B \: ; \: m_{\Ap} \big)} & \hat{\epsilon} \leq \epsilon \\
    0 & \hat{\epsilon} > \epsilon 
    \end{cases}
    \label{eq:q_eps}
\end{equation}
where $\epsilon'$ is the value of $\epsilon$ under which the pseudo-data $\vec{d}$ are generated (with $\vec{d}(0)$ corresponding to data generated under the background only hypothesis), $(\hat{\epsilon}, \hat{\theta}_S, \hat{\theta}_B)$ are the unconditional maximum likelihood estimators, and $(\hat{\hat{\theta}}_S, \hat{\hat{\theta}}_B)$ are the conditional maximum likelihood estimators of the nuisance parameters at fixed $\epsilon$. The test statistic is set to zero when $\hat{\epsilon} > \epsilon$ to enforce the one-sided nature of the test: data that prefer a signal strength 
larger than the tested value are not taken as evidence against that hypothesis.

The asymptotic approximations of the test statistic distributions derived in \cite{Likelihoods}, which follow from Wilks' theorem, are valid only in the large-sample limit. In this regime, the profile likelihood ratio test statistic converges to a half-$\chi^2$ distribution, substantially reducing the computational cost of evaluating $p$-values. However, given the low expected bin counts in both the signal and background distributions, this approximation is not reliable here. 
We therefore evaluate the $p$-value by sampling the distributions of $q_\epsilon$ directly from toy Monte Carlo pseudo-experiments, generating pseudo-data under both the signal-plus-background and background-only hypotheses for each value of $\epsilon$ and $m_{\Ap}$.

\subsubsection{Discovery}
\label{section:Discovery_Methods}
To claim a discovery, we must reject the background only hypothesis at a given significance. We consider a significance of $5 \sigma$ to be \textit{discoverable}, with the significance defined as \mbox{$Z \equiv \Phi^{-1}(1-p)$}, where $\Phi^{-1}$ is the inverse of the standard normal cumulative distribution function and the $p$-value is,
\begin{equation}
    p = \int_{\text{med}[f(q_0|\epsilon)]}^{\infty} f(q_0|0)\, dq_0,
\end{equation}
where $f(q_0|0)$ is the distribution of the test statistic with data simulated under the background-only hypothesis and $\text{med}[f(q_0|\epsilon)]$ is the median of the test statistic distribution with data simulated under signal-plus-background with $\epsilon$. Similarly to the exclusion case, the median of $f(q_0|\epsilon)$ replaces the observed $q_0$ since there is no data yet -- providing a median significance and thus quantifying the \textit{expected significance}. 
 The expected discovery significance of $5\sigma$ at a given $\epsilon$ and $m_{A'}$ means that, if a signal of that strength exists, half of all pseudo-experiments would yield a significance exceeding $5\sigma$.

The test statistic for discovery is \cite{Likelihoods},
\begin{equation}
    q_0 =
    \begin{cases}
    -2 \log \frac{\mathcal{L}\big(\vec{d}(\epsilon') \: | \: \epsilon=0, \: \hat{\hat{\theta}}_S, \: \hat{\hat{\theta}}_B \: ; \: m_{\Ap} \big)}{\mathcal{L}\big(\vec{d}(\epsilon') \:|\:\hat{\epsilon}, \:\hat{\theta}_S, \: \hat{\theta}_B \: ; \: m_{\Ap} \big)} & \hat{\epsilon} \geq 0 \\
    0 & \hat{\epsilon} < 0,
    \end{cases}
    \label{eq:q0}
\end{equation}
where the definitions of the variables are given under \eqnref{eq:q_eps}. Notice that the test statistic $q_0$ is set to 0 when $\hat{\epsilon} < 0$, since a deficit of events relative to background is not considered evidence for a signal.

Similarly to the procedure for projected exclusion bounds, the asymptotic approximations of the test statistic distributions from \cite{Likelihoods} are not always valid within this analysis. We therefore sample both of the test statistic distributions $f(q_0|0)$ and $f(q_0|\epsilon)$ directly via pseudo-experiments, ensuring that the resulting discovery reach is accurate even in regimes where the asymptotic approximations break down. Due to increased computational cost with an increased discovery significance, for cases yielding $> 5\sigma$ sensitivity we use the asymptotic approximations as described in \cite{Likelihoods}.

The discovery significances reported here are local values, computed with the benchmark mediator mass $m_{A'}$ held fixed. In a realistic search in which $m_{A'}$ is unknown and must be scanned over, the corresponding global significance would be reduced relative to these local values owing to the trial factor incurred by testing multiple mass hypotheses.

\subsubsection{Parameter Estimation}
\label{section:ParameterEstimation}

With a sufficient number of signal events, LDMX will not only have the ability to claim a discovery, but could also infer the mass and coupling of the underlying dark photon model.
To estimate the parameters $(m_{A'}, \epsilon)$ and their uncertainties, we perform a scan of
the profile likelihood ratio,
\begin{equation} \Delta(-2\log \mathcal{L}) = -2\log  \frac{\mathcal{L}(\vec{d}(\epsilon')\,|\,m_{A'},\epsilon, \hat{\hat{\theta}}_S,\hat{\hat{\theta}}_B)}{\mathcal{L}(\vec{d}(\epsilon')\,|\,\hat{m}_{A'}, \: \hat{\epsilon}, \:\hat{\theta}_S, \: \hat{\theta}_B)},
    \label{eq:profile_llr}
\end{equation}
where the definitions of the variables are given under \eqnref{eq:q_eps}. Note that in this expression, $m_{A'}$ is not longer fixed, but varies with the scan along with $\epsilon$.

For the sake of drawing $1\sigma$ and $2\sigma$ confidence regions, we assume Wilks' theorem holds. Therefore, $\Delta(-2\log L)$  is asymptotically $\chi^2$-distributed with $k$ degrees of freedom, where $k$
is the number of parameters of interest, 2. For a joint scan over
$(m_{A'}, \epsilon)$, the $1\sigma$ (68\%) and $2\sigma$ (95\%) confidence regions are bounded by
$\Delta(-2\log L) < 2.30$ and $\Delta(-2\log L) < 6.18$.

As a proxy for the precision and accuracy of the maximum likelihood parameter estimation of $\epsilon$ and $m_{A'}$, we compute the following quantities.

To characterize the statistical performance of the maximum likelihood parameter estimation of $\epsilon$ and $m_{A'}$, we report two quantities for each benchmark point, computed from an ensemble of 25 pseudo-experiments, the relative bias $b_X$, and the relative half-interval width $\sigma_X$,
\begin{equation}
    b_X \equiv \frac{\hat{X}-X_{\text{true}}}{X_{\text{true}}}, \: \sigma_X \equiv \frac{X_+ - X_-}{2X_{\text{true}}}
    \label{eq:bias_width}
\end{equation}
where $\hat{X}$ is the median best-fit value across the ensemble and $X_\pm$ are the boundaries of the $1\sigma$ confidence interval under the asymptotic $\chi^2$ approximation. We report the absolute value of the bias $b_X$. Here, $X$ represents the parameters of interest, namely $m_{A'}$ or $\epsilon$. The bias measures how accurately the estimator recovers the true parameter, and the half-interval width measures the statistical precision of the inference.
Both quantities are normalized by the true parameter value, placing them on a common scale and making them directly comparable across benchmark points where $m_{A'}$ and $\epsilon$ vary by orders of magnitude. We report the median of each quantity over the ensemble of pseudo-experiments, as the small sample size renders individual realizations sensitive to statistical fluctuations.

\subsection{Bayesian Statistical Framework}
\label{sec:bayesian}

The Bayesian framework is built upon the likelihood introduced at the beginning of this section,
with the key distinction that the nuisance parameters, $\theta_S$ and $\theta_B$, are now marginalised over with dedicated priors rather than profiled out. Their priors are the log-normal distributions defined as constraint terms in \eqnref{eq:likelihood}. 
Rather than fixing the width of the background nuisance prior, $\sigma_B$, to a single value, we treat it as an unknown hyperparameter and place a hyperprior on it. This hierarchical construction
allows the data themselves to inform the degree of background uncertainty, avoiding the need to assume a precise value for $\sigma_B$ a priori. We adopt a half-normal hyperprior,
\begin{equation}
    \pi(\sigma_B) = {\frac{\sqrt2}{\sigma \sqrt\pi }} \exp\left( -\frac{\sigma_B^2}{2\sigma^2} \right),
\end{equation}
whose shape parameter, $\sigma = 0.5$.

The parameters $m_{A'}$ and $\epsilon$ both span several orders of magnitude over their physically
motivated ranges, making a log-uniform prior the natural choice. The log-uniform prior reflects ignorance of the order of magnitude of these couplings and ensures that the inference is not artificially biased toward any particular decade of parameter space. The priors are defined over the range $m_{A'} \in \big[ 10^{-3}, 1 \big]$ GeV and $\epsilon \in \big[ 10^{-20}, 1.0 \big]$. The analogous priors are defined for the additional coupling constants entering the DEM models of \eqnref{eq:lagrangian}. 

The joint posterior over all parameters of interest and nuisance
parameters is sampled via Bayes' theorem,
\begin{equation}
    p(m_{A'}, \epsilon, \theta_S, \theta_B, \sigma_B \mid \vec{d}\,) \propto \mathcal{L}(\vec{d}\mid m_{A'}, \epsilon, \theta_S, \theta_B)\,\pi(m_{A'}, \epsilon, \theta_S, \theta_B,\sigma_B),
\end{equation}
where $\mathcal{L}$ is the binned two-dimensional Poisson likelihood of \eqnref{eq:likelihood}
and $\pi$ is the joint prior over all parameters, which is simply the product of each parameter's prior.

The joint posterior over $\{m_{A'},\,\epsilon,\,\theta_S,\,\theta_B,\,\sigma_B\}$ is explored using dynamic
nested sampling as implemented in \texttt{dynesty}~\cite{Speagle:2019ivv}.  
In contrast to standard (static) nested sampling, which propagates a fixed number of live points throughout the run, dynamic nested sampling adaptively allocates additional live points during the run to the regions of parameter space that contribute most \cite{Higson:2018cwj}. This targeted allocation improves sampling efficiency over the static algorithm in settings such as ours, where both well-resolved posteriors for parameter estimation and precise evidence estimates for Bayes factor calculations are required. Nested sampling is particularly well-suited to this analysis: it handles the multi-dimensional, multi-modal posterior efficiently, and simultaneously provides a direct estimate of the Bayesian evidence used for model comparison, described below.
The resulting posterior distributions of the parameters provide an estimate of the uncertainty of the inferred parameter values. We use generated pseudo-data at the benchmark parameter values of \figref{fig:BMs} to determine the projected uncertainties (the $1\sigma$ credible regions), and the relative bias from \eqnref{eq:bias_width} of a hypothetical parameter inference in the case of a positive signal. We report the median values of the $1\sigma$ credible regions and relative biases from ten pseudo-experiments and their corresponding Bayesian posterior scans.

\subsubsection{Model Comparison}
\label{sec:model_comparison}

A key advantage of the Bayesian framework is that posterior inference naturally yields the Bayesian
evidence,
\begin{equation}
    \mathcal{Z} = \int \mathcal{L}(\vec{d}\:|\:\vec{\nu})\,\pi(\vec{\nu})\,\mathrm{d}\vec{\nu},
\end{equation}
where $\vec{\nu} \equiv (g_f,\,m_{A'},\,\theta_S,\,\theta_B,\,\sigma_B)$ and $g_f$ is $\epsilon$, $\mu_f/e$, $d_f/e$, $b_f/e$, or $a_f/e$, for the different dark photon models. The Bayesian evidence is used to perform model comparison between competing signal hypotheses. Given two models $\mathcal{M}_1$ and $\mathcal{M}_2$, the Bayes factor is defined as the ratio of their respective evidences,
\begin{equation}
   K_{12} = \frac{\mathcal{Z}_1}{\mathcal{Z}_2}
    = \frac{\int \mathcal{L}(\vec{d}\mid\vec{\nu}_{\mathcal{M}_1})\,\pi(\vec{\nu}_{\mathcal{M}_1})\,\mathrm{d}\vec{\nu}_{\mathcal{M}_1}}
    {\int \mathcal{L}(\vec{d}\mid\vec{\nu}_{\mathcal{M}_2})\,\pi(\vec{\nu}_{\mathcal{M}_2})\,\mathrm{d}\vec{\nu}_{\mathcal{M}_2}},
\end{equation}
where $\vec{\nu}_{\mathcal{M}}$ is the set of inferred parameters for model $\mathcal{M}$. We drop the indices 12 and refer to $K_{12}$ as simply $K$ for the remainder of the paper.
We compute $K$ for each pair of dark photon models considered in this work. The Bayes factor thereby quantifies how much more strongly the full
kinematic information supports one model over another, providing a direct answer to the question of whether LDMX can distinguish between different dark photon signal hypotheses at a given set of model
parameters. We take the median Bayes factor from a set of five pseudo-experiments.

The strength of evidence is assessed using the Jeffreys scale, which provides a standard interpretive framework for the Bayes factor \cite{Jeffreys:1939xee,Berger:2001zbn}. Values of the Bayes factor above the decisive threshold indicate that the kinematic distributions carry sufficient information to discriminate between the models at the given signal parameters, while values below the substantial threshold suggest the models are effectively indistinguishable with the current analysis. Due to the statistical fluctuation of the \madgraph generated signal modelling, we consider a stricter criteria than the standard $K > 100$ for decisive evidence against the alternative hypothesis $\mathcal{M}_2$ (distinguishable), where we require $\log K \gtrsim 10$. 

Certain dark photon model hypotheses are inherently indistinguishable due to their recoil $e^-$ spectra having the same shape. The overall normalization, however, can vary between these spectra of the same shape. To exploit the difference that does exist in the signal normalization of these model pairs, we add a prior on $g_f$ that penalizes fit values far from the relic density curve. The prior $\pi(g_f)$ is a normal distribution centred with its median on the $g_f$ values of Fig. 1 of \cite{Catena:2025fsl}, with standard deviation $\sigma_{g_f} = 0.1$.

\section{Frequentist Analysis}
\label{section:frequentist}
In this section, we present the results of the frequentist analysis using the likelihood framework described in \secref{section:Likelihood}.

\subsection{Projected Exclusion Bounds}

\begin{figure}
    \centering
\includegraphics[width=0.9\linewidth]{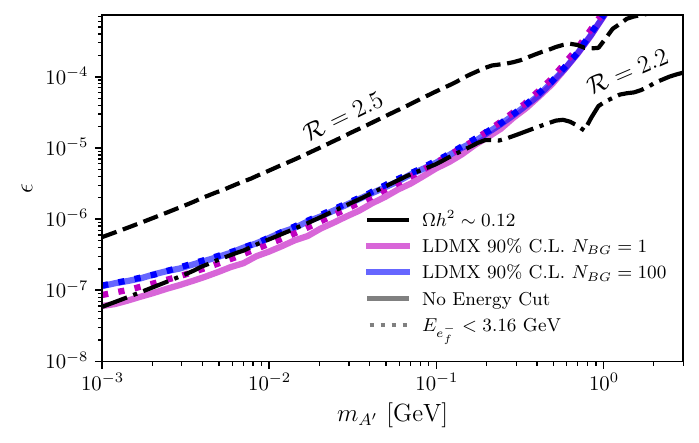}
    \caption{The projected $90\%$ C.L. exclusion bound on $\epsilon$ as a function of $m_{A'}$ from our frequentist likelihood framework, in magenta (blue) with 1 (100) background events. The dotted curves apply an energy cut of $E_{e^-_f} < 3.16$ GeV, while the solid curves apply no energy cuts. The relic targets for $\mathcal{R} = 2.5$ and $2.2$ are drawn in dashed and dash-dotted, respectively.}
    \label{fig:exclusion}
\end{figure}

The projected $90\%$ C.L. exclusion bounds on the kinetic mixing parameter $\epsilon$ as a function of the dark photon mass $m_{A'}$ are shown in \figref{fig:exclusion}. These are obtained from our frequentist likelihood framework incorporating the full two-dimensional kinematic distributions together with systematic uncertainties on the signal and background.
The $90\%$ exclusion bound is the contour in the $(\epsilon,\, m_{A'})$ plane at which the exclusion $p$-value equals $0.1$. Excluding a point on the relic target corresponds to ruling 
out thermal freeze-out production of dark photon-mediated DM at that mass ratio $\mathcal{R}$. Our analysis projects that, in the case of a null result, the relic target is excluded for $\mathcal{R} \gtrsim 2.2$ across the mass range $m_{A'} \in [10^{-3},\, 0.2]$~GeV, with the precise reach depending on $\mathcal{R}$ and the background level. As expected, an increased background rate reduces the sensitivity. Notice that for small masses, the two background scenarios differ most, however, for larger masses, the exclusion sensitivity loses dependence on the background. This is because at higher masses, the signal occupies more of the high $|p_T|_{e^-_f}$ bins which have no expected background.

Applying the energy cut $E_{e^-_f} < 3.16$~GeV has a number of effects on the exclusion limits. First, it reduces not only the background, but also the signal acceptance. Second, the factor of $\sim 10$ reduction in the background weakens the constraint on the background normalization nuisance parameter $\theta_B$. A less tightly constrained $\theta_B$ increases the degeneracy between signal strength and background normalization in the profile likelihood, allowing a genuine signal-like excess to be partially reabsorbed into the background component rather than attributed to signal, which offsets some of the sensitivity gain expected from the improved signal to background ratio in the retained bins. Third, part of the signal shape that is used for discriminating signal from background is lost. For $N_\mathrm{BG}=1$, where the background normalization constraint is already relatively weak in the no-cut case, this further degradation outweighs the signal to background effect, and the exclusion limit worsens under the cut. For $N_\mathrm{BG}=100$, the background normalization remains comparatively well constrained in both the cut and no-cut configurations, so the loss in shape information and signal acceptance from the cut is approximately compensated by the improvement in signal to background, and the two exclusion curves are nearly coincident.

Comparing the $N_\text{BG}=1$ results to a single-bin analysis within the same setup (with systematic uncertainties and background), the single-bin results are less sensitive by a factor of approximately 1.3 in $\epsilon$, indicating that the shape information provides additional discriminating power between the signal and background hypotheses.

Conventionally, collider based searches often apply the
$\mathrm{CL}_s$ method~\cite{Read:2002hq}, 
which guards against spurious exclusions of signal hypotheses in regions of low sensitivity or in the presence of downward background fluctuations, by normalizing the $p$-value of the signal-plus-background hypothesis to that of the background-only hypothesis. We therefore additionally compute the exclusion bounds applying this modification. We find that the $\mathrm{CL}_s$ exclusion bound 
is more conservative than the standard bound by a factor of ${\simeq}\,1.2$ across the parameter space.

\subsection{Projected Discovery Significance}
\label{sec:discovery}

We evaluate the projected discovery sensitivity of LDMX Phase II at four benchmark points along the relic density target, following the frequentist discovery procedure described in \secref{section:Discovery_Methods}. The median expected significance is computed from the distribution of the discovery test statistic
$q_0$ defined in \eqnref{eq:q0} evaluated via pseudo-experiments. As discussed in \secref{section:Discovery_Methods}, the asymptotic approximation is not reliable in the
regime of low expected bin counts; accordingly, the $p$-value is derived by sampling
$f(q_0|0)$ and $f(q_0|\epsilon)$ directly from pseudo-experiments for all benchmark points
where the resulting significance is near or below $5\sigma$. For cases yielding significance
well in excess of $5\sigma$, the asymptotic approximation of Ref.~\cite{Likelihoods} is used
and the result is quoted as $\gg 5$.
\renewcommand{\arraystretch}{1.2}
\begin{table}[h]
\centering
\caption{Expected discovery sensitivity at four benchmark points along the relic target. $S$ and $B$ denote the expected signal and background counts respectively. The median expected discovery significance $Z$ is derived from full pseudo-experiments, except for $Z \gg 5$ where we rely on the asymptotic approximation. $Z \geq 5$ is classified as~\textit{discoverable}. $Z_\text{cut}$ refers to an applied energy cut $E_{e^-_f} < 3.16$ GeV.}
\label{tab:discovery}
\begin{tabular}{lcccccccc}
\toprule \toprule
& $m_{\Ap}$ [GeV] & $\mathcal{R}\equiv \frac{m_{\Ap}}{m_\text{DM}}$ & $\epsilon$ & $S$ & $Z_{B=1}$ & $Z_{B=100}$ & $Z^\text{cut}_{B=1}$ & $Z^\text{cut}_{B=100}$ \\
\hline
\textbf{BM 1} & 0.01 & 2.5 & 4.8 $\times 10^{-6}$ & 310 & $\gg 5$ & $\gg 5$ & $\gg 5$ & $\gg 5$ \\
\textbf{BM 2} & 0.01 & 2.2 & 5.2 $\times 10^{-7}$ & 4   & 4.4  &  1.3 & 4.3 & 1.3 \\
\textbf{BM 3} & 0.1  & 2.5 & 6.3 $\times 10^{-5}$ & 250 & $\gg 5$ & $\gg 5$ & $\gg 5$ & $\gg 5$ \\
\textbf{BM 4} & 0.1  & 2.2 & 6.0 $\times 10^{-6}$ & 2   & 4.2  & 2.2 & 3.9 & 1.7 \\
\toprule \toprule
\end{tabular}
\end{table}

The results are summarized in Table~\ref{tab:discovery} for two background scenarios, $N_\text{BG} = 1$ and $N_\text{BG} = 100$, and both with and without the recoil electron energy cut $E_{e_f^-} \leq 3.16\,\mathrm{GeV}$ recommended for
Phase II. The distributions of $q_0$ under the background-only and signal-plus-background hypotheses for each benchmark are shown in \appref{section:test_statistic_dists}.
For benchmark points 1 and 3, which lie on the $\mathcal{R} = 2.5$ relic target and carry larger signal yields, the projected significance
is far in excess of $5\sigma$ in both background scenarios and regardless of whether the
energy cut is applied. In these cases, the two $q_0$ distributions are well separated with
negligible overlap, as visible in Figs.~\ref{fig:discovery_q} and \ref{fig:discovery_q_Ecut}, and a discovery is
effectively guaranteed if the signal exists: benchmark 1 and 3 are discoverable.

Benchmark points 2 and 4, situated on the more challenging $\mathcal{R} = 2.2$ relic target with correspondingly smaller signal yields, probe the threshold of the discovery reach. In the low background scenario ($N_\text{BG} = 1$), both benchmarks yield a median significance just below $5\sigma$, indicating that discovery at the $5\sigma$ level is just out of reach. In the high background scenario ($N_\text{BG} = 100$), the sensitivity is significantly
degraded.
This demonstrates that controlling the background level is critical for achieving sensitivity along the $\mathcal{R} = 2.2$ relic target.

The energy cut further reduces the discovery reach only slightly at the $m_{A'}=0.1$ GeV $\mathcal{R} = 2.2$ benchmark, however for the $m_{A'}=0.01$ GeV $\mathcal{R} = 2.2$ benchmark the discovery significance is almost equal.
By contrast, the $\mathcal{R} = 2.5$ benchmarks remain unaffected, as their high signal yields keep the significance well above $5\sigma$ even after
the cut is applied.

Taken together, these results demonstrate that LDMX Phase II has strong projected discovery potential along the $\mathcal{R} = 2.5$ relic target across the full range of background scenarios considered. Discovery along the $\mathcal{R} = 2.2$ target is not quite achievable at $5
\sigma$.

\subsection{Maximum Likelihood Parameter Estimation}
\label{sec:mle}

We assess the ability of LDMX to infer the dark photon parameters
$(\epsilon,\, m_{A'})$ in the event of a signal discovery, using the profile likelihood ratio scan defined in \secref{section:ParameterEstimation}. For each benchmark point, 25
pseudo-data sets are generated at the true parameter values and the profile likelihood ratio test statistic from \eqnref{eq:profile_llr} is computed over the $(\epsilon,\, m_{A'})$ plane for each. The resulting profile likelihood ratios are averaged across the ensemble prior to visualization, since the low signal yields at several benchmark points render individual pseudo-data realizations susceptible to large statistical fluctuations; a single pseudo-data set would therefore be an unreliable visual representation of the typical inference.

\begin{figure}[H]
    \centering
\includegraphics[width=0.72\linewidth]{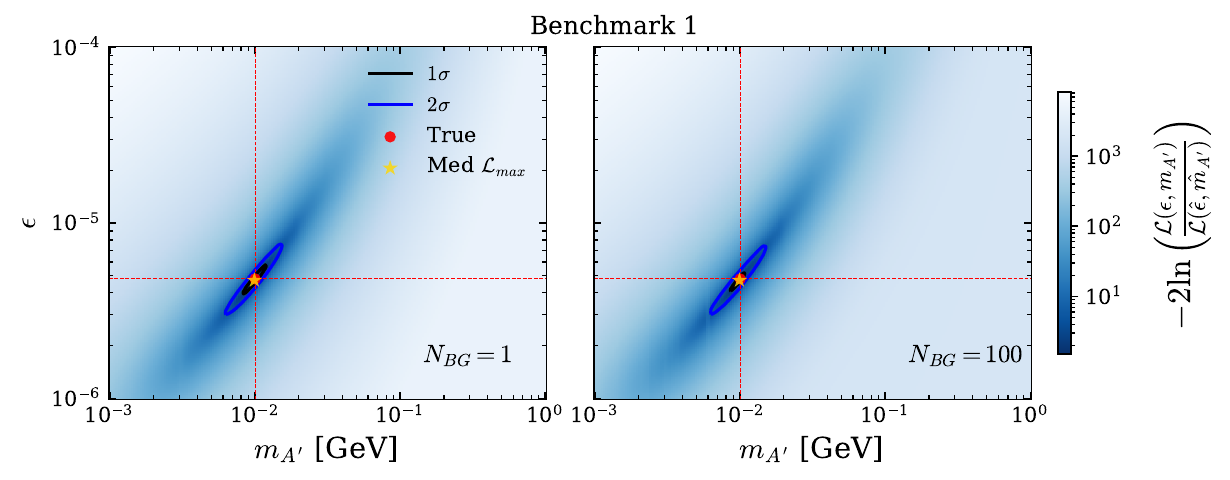}
\includegraphics[width=0.72\linewidth]{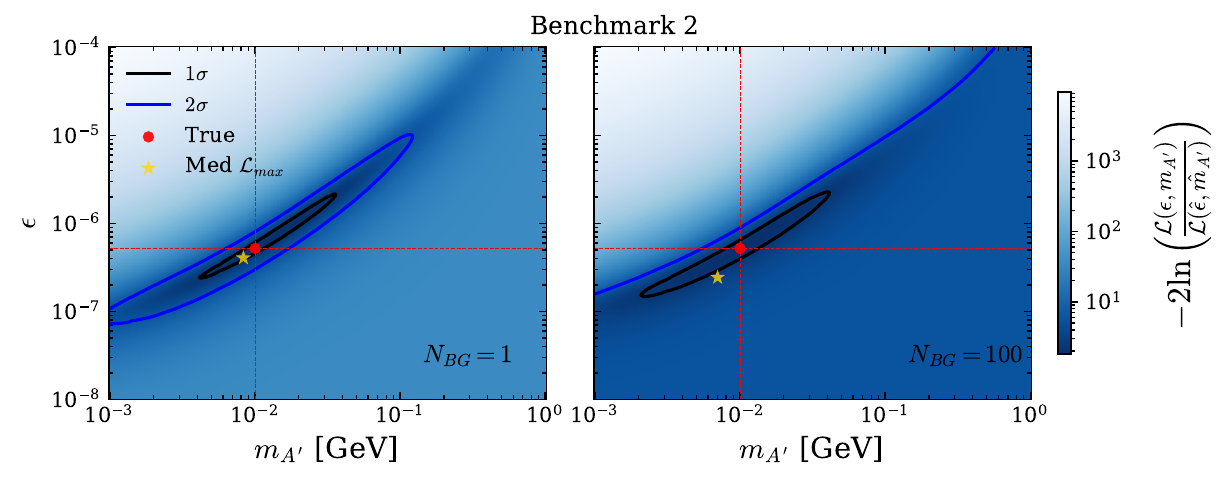}
\includegraphics[width=0.72\linewidth]{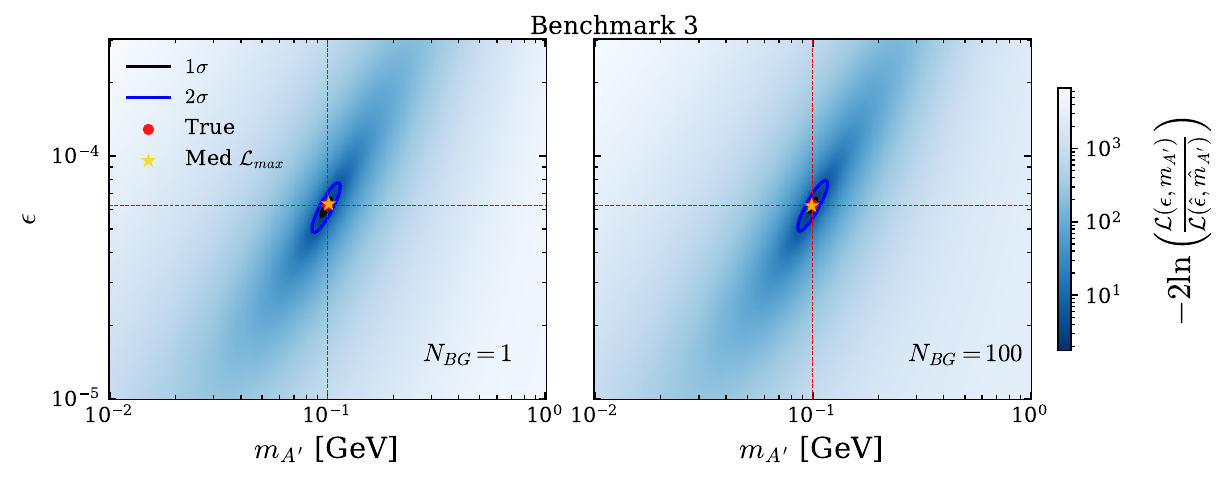}
\includegraphics[width=0.72\linewidth]{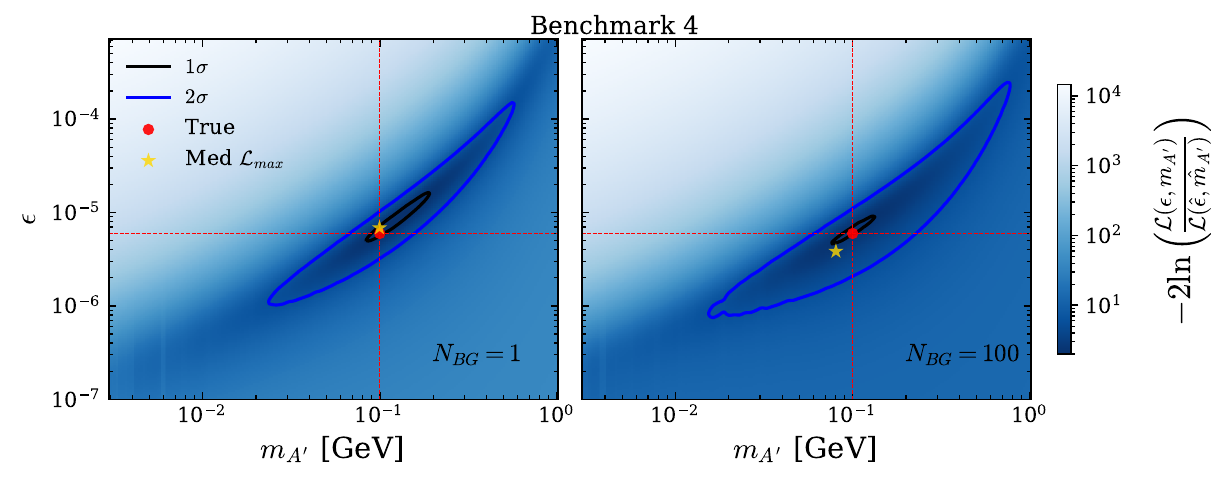}
    \caption{
    The profile likelihood ratio
    in the $(\epsilon,\, m_{A'})$ plane for each of the four benchmark points (rows),
    evaluated for background scenarios $N_\text{BG} = 1$ (left) and $N_\text{BG} = 100$ (right).
    The profile likelihood ratio is averaged over 25 pseudo-data sets generated at the true parameter values, indicated by the red points. The yellow star marks the median maximum likelihood estimate $(\hat{\epsilon},\, \hat{m}_{A'})$ across
    pseudo-experiments. The black (blue) contours denote the $1\sigma$ ($2\sigma$)
    confidence regions, defined under the asymptotic $\chi^2$ approximation with $k = 2$
    degrees of freedom as the set of parameter points satisfying
    $\Delta(-2\log\mathcal{L}) < 2.30$ ($6.18$).}
\label{fig:profile_likelihood}
\end{figure}

The resulting contour maps are shown in
\figref{fig:profile_likelihood} for both background scenarios, $N_\text{BG} = 1$ and $N_\text{BG} = 100$. The $1\sigma$ ($2\sigma$) confidence regions are defined as the sets of parameter points satisfying ${\Delta(-2\log\mathcal{L}) < 2.30}$ ($6.18$) under the asymptotic $\chi^2$ approximation with $k = 2$ degrees of freedom.
To quantify the accuracy and precision of the parameter inference more concisely, we report the median relative bias $b_X$ and the median relative half-interval width $\sigma_X$ from \eqnref{eq:bias_width}
for each parameter $X \in \{m_{A'},\, \epsilon\}$, computed over the ensemble of 25 pseudo-experiments, summarized in \tabref{tab:ML}.
\renewcommand{\arraystretch}{1.2}
\begin{table}[h]
\centering
\caption{Maximum profile likelihood parameter estimation uncertainties for four benchmark points. The uncertainty quantities reported are the relative bias ($b$) and the relative half-interval width ($\sigma$) defined in \eqnref{eq:bias_width} for two background scenarios. $\sigma_X$ entries marked with a -- are uninformative, with half-width values reaching the border of the prior.}
\label{tab:param_estimate_benchmarks}
\begin{tabular}{l c c cc cc}
\toprule \toprule
& & & \multicolumn{2}{c}{\textbf{Background = 1}} & \multicolumn{2}{c}{\textbf{Background = 100}} \\
\cmidrule(lr){4-5} \cmidrule(lr){6-7}
& $m_{\Ap}$ [GeV] & $\epsilon$ & $b_{m_{\Ap}}$ ($\sigma_{m_{\Ap}}$) & $b_{\epsilon}$ ($\sigma_{\epsilon}$) & $b_{m_{\Ap}}$ ($\sigma_{m_{\Ap}}$) & $b_{\epsilon}$ ($\sigma_{\epsilon}$) \\ 
\midrule
\textbf{BM 1} & 0.01 & $4.8 \times 10^{-6}$ & 0.1 (0.2) & 0.2 (0.2) & 0.1 (0.2) & 0.1 (0.2) \\
\textbf{BM 2} & 0.01 & $5.2 \times 10^{-7}$ & 0.4 (10)  & 0.3 (3) & 0.6 (--)   & 0.18 (3) \\
\textbf{BM 3} & 0.1  & $6.3 \times 10^{-5}$ & 0.01 (0.07) & 0.007 (0.1) & 0.01 (0.07)   & 0.007 (0.1) \\
\textbf{BM 4} & 0.1  & $6.0 \times 10^{-6}$ & 0.4 (0.6)   & 0.6 (3) & 0.6 (0.9)    & 0.9 (2) \\
\bottomrule \bottomrule
\label{tab:ML}
\end{tabular}
\end{table}

For benchmark points 1 and 3, situated on the $\mathcal{R} = 2.5$ relic target with larger signal yields, both parameters are recovered with relatively high accuracy and precision in both background scenarios.
This behaviour is also reflected in \figref{fig:profile_likelihood}, where the profile likelihood ratio contours for benchmarks 1 and 3 form tight, well localized ellipses centred
close to the true parameter values (red point). Notice that a higher background level at these benchmarks do not increase the uncertainty on the inferred parameters.

Benchmark points 2 and 4, on the $\mathcal{R} = 2.2$ relic target, present a more challenging inference problem owing to their small signal yields. The 1 and 2$\sigma$ regions along with the maximum likelihood point vary quite drastically across pseudo-data samples.
With signal yields of $S = 5$ and $S = 3$, individual pseudo-data realizations at these benchmarks are dominated by Poisson fluctuations, producing contour shapes that differ substantially from one experiment to the next.
This leads to a larger variance in the reconstructed parameters, and thus larger variances in the biases and half-interval widths.
Even though the uncertainties presented in \tabref{tab:ML} may not be so large, the large variance between pseudo-data samples leads to an unreliable inference.
These larger variances in the uncertainties are not reported in \tabref{tab:ML}, however they are $\sim$ 1 to 2 orders of magnitude larger than the variances of benchmark 1 and 3 uncertainties.

For benchmark 2, this manifests in the averaged profile likelihood as an open $2\sigma$ contour --- no closed loop exists at the $2\sigma$ level in one or more directions of the $(\epsilon,\, m_{A'})$ plane. This indicates that a large region of parameter space remains statistically consistent with the data at the $2\sigma$ level, reflecting the inability of a $S = 5$ event signal to meaningfully discriminate between a wide range of $(\epsilon,\, m_{A'})$ combinations that produce similarly small expected yields.

For benchmark 4, the averaged $1\sigma$ contour appears deceptively tight. We interpret this as an artifact of the averaging procedure. When the profile likelihood from 25 pseudo-experiments, each individually broad and centred at a different location due to statistical fluctuations, are superimposed, their average can exhibit a spuriously
narrow peak near the true parameter values if the individual minima are approximately
symmetrically scattered around the truth. The resulting averaged contour then understates the uncertainty that any single experimental realization would actually report. In addition, the perceived improvement of the uncertainties from 1 to 100 background events at benchmark 4 is not physical, but merely the similar effect of averaging described above.

Overall, these results indicate that LDMX can achieve precise and accurate
parameter estimation along the $\mathcal{R} = 2.5$ relic target for both background
scenarios considered. For the $\mathcal{R} = 2.2$ target where the signal yield is small, the inference quality is more sensitive to the statistical fluctuations, therefore a higher luminosity would be required for reliable parameter inference.

\section{Bayesian Analysis}
\label{section:bayesian}
The results for the Bayesian analysis, with the methodology described in \secref{sec:bayesian} are presented in this section.

\subsection{Posterior Distributions and Credible Regions}
\begin{table}[h]
\centering
\caption{Bayesian parameter estimates and credible regions for the dark photon mass $m_{A'}$ and kinetic mixing parameter $\epsilon$ at each benchmark point, recovered using dynamic nested sampling under background scenarios $N_\mathrm{BG} = 1$ and $N_\mathrm{BG} = 100$. Quoted values are median quantities from 10 random pseudo-data sets, where for each pseudo-data set the medians of the posterior distributions over the model parameters and their asymmetric 68\% credible intervals derived from the 16th and 84th percentiles are computed.
The corresponding parameter inferences with the $E_{e_f^-} < 3.16$ GeV cut are included. Entries with dashes correspond to uninformative inferences, where the posterior credible interval spans the full prior range.}
\centering
\setlength{\tabcolsep}{2pt}
\begin{tabular}{cccc c c c c}
\hline\hline
 & $m_{A'}$ [GeV] & $\epsilon$ & BG & $\hat{m}_{A'}$ [GeV] & $\hat{\epsilon}$ & $\hat{m}_{A',\text{cut}}$ [GeV] & $\hat{\epsilon}_{\text{cut}}$ \\
\hline
\multirow{2}{*}{\textbf{BM 1}} & \multirow{2}{*}{0.01} & \multirow{2}{*}{$4.8\times10^{-6}$}
    & 1   & $0.011^{+0.003}_{-0.002}$ & $(6^{+1}_{-1})\times10^{-6}$ & $0.012^{+0.004}_{-0.003}$ & $(6^{+2}_{-1})\times10^{-6}$ \\
  & & & 100 & $0.011^{+0.003}_{-0.002}$ & $(5^{+2}_{-1})\times10^{-6}$ & $0.011^{+0.004}_{-0.003}$ & $(5^{+2}_{-1})\times10^{-6}$ \\
\hline
\multirow{2}{*}{\textbf{BM 2}} & \multirow{2}{*}{0.01} & \multirow{2}{*}{$5.2\times10^{-7}$}
    & 1   & $0.02^{+0.10}_{-0.02}$ & $(0.5^{+2.5}_{-0.4})\times10^{-6}$ & $0.02^{+0.09}_{-0.01}$ & $(0.4^{+2.6}_{-0.3})\times10^{-6}$ \\
  & & & 100 & $0.03^{+0.31}_{-0.03}$ & -- & $0.03^{+0.32}_{-0.03}$ & -- \\
\hline
\multirow{2}{*}{\textbf{BM 3}} & \multirow{2}{*}{0.10} & \multirow{2}{*}{$6.3\times10^{-5}$}
    & 1   & $0.099^{+0.009}_{-0.007}$ & $(6.3^{+0.8}_{-0.8})\times10^{-5}$ & $0.096^{+0.009}_{-0.009}$ & $(6.1^{+0.9}_{-0.8})\times10^{-5}$ \\
  & & & 100 & $0.096^{+0.005}_{-0.007}$ & $(6.1^{+0.6}_{-0.7})\times10^{-5}$ & $0.097^{+0.008}_{-0.008}$ & $(6.1^{+0.8}_{-0.8})\times10^{-5}$ \\
\hline
\multirow{2}{*}{\textbf{BM 4}} & \multirow{2}{*}{0.10} & \multirow{2}{*}{$6.0\times10^{-6}$}
    & 1   & $0.04^{+0.29}_{-0.04}$ & $(0.5^{+5.6}_{-0.4})\times10^{-6}$ & $0.04^{+0.28}_{-0.03}$ & $(0.4^{+4.8}_{-0.3})\times10^{-6}$ \\
  & & & 100 & $0.06^{+0.30}_{-0.05}$ & -- & $0.06^{+0.29}_{-0.05}$ & -- \\
\hline\hline
\end{tabular}
\label{tab:param_estimate_benchmarks}
\end{table}

Table~\ref{tab:param_estimate_benchmarks} summarizes the median Bayesian parameter estimates for $m_{A'}$ and $\epsilon$ at each benchmark point, quoted as posterior medians with asymmetric 68\% credible intervals over 10 pseudo-data sets. At benchmark 1 and 3, both parameters are recovered within the uncertainties, with posterior medians consistent with the injected values to within $1\sigma$ across both background scenarios. The credible intervals at benchmark 3 are notably 
more symmetric than at benchmark 1, reflecting the higher number of expected signal events and 
correspondingly more Gaussian posterior shape. At benchmark 2 and 4, the credible intervals are large and prior dependent, especially in the lower limits ($-\sigma$). The $E_{e^-_f} < 3.16$ GeV cut analysis shows no significant difference between the un-cut analysis.

\begin{figure}
    \centering
\includegraphics[width=\linewidth]{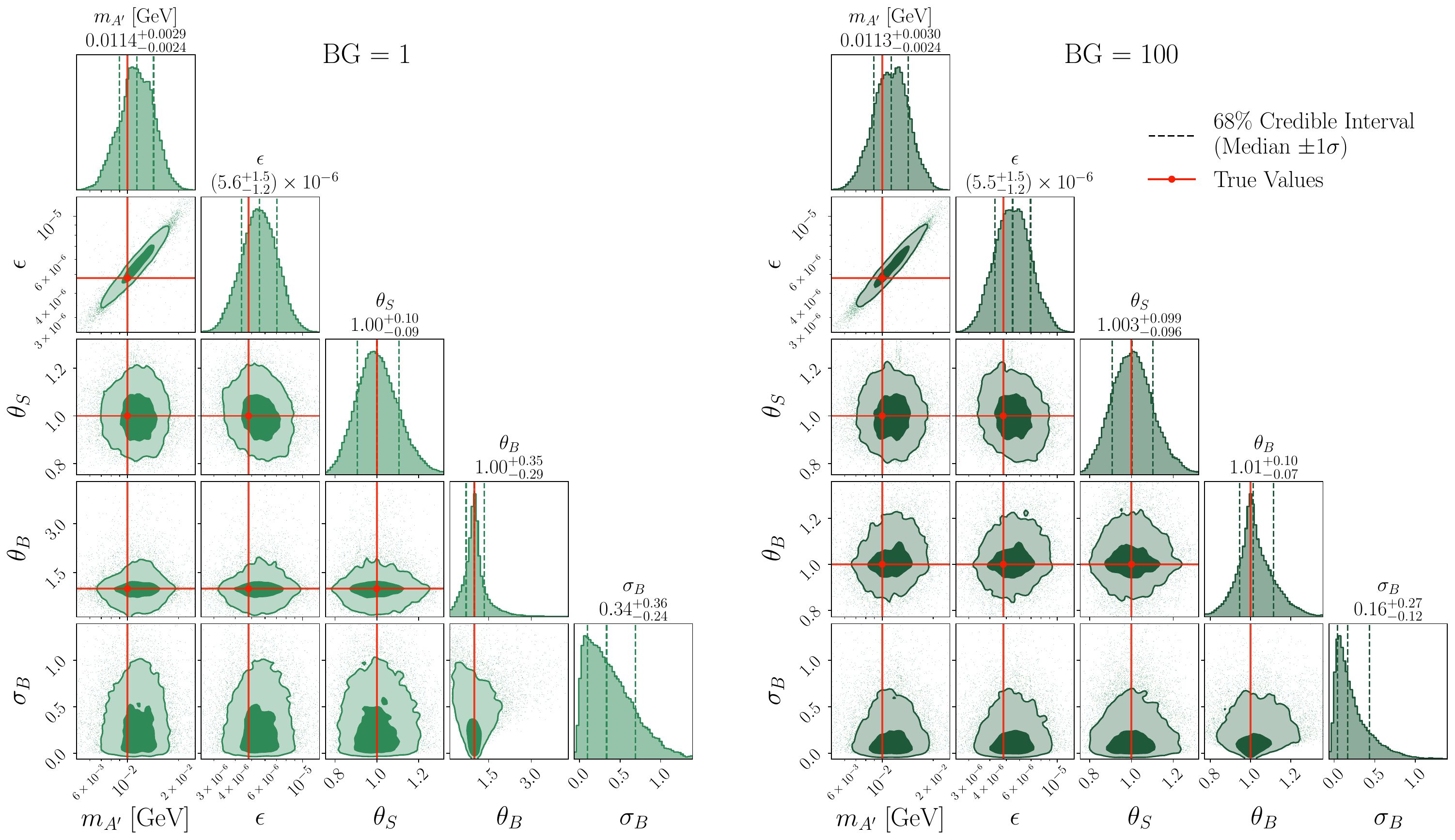}
\caption{Corner plots of the marginalized posterior distributions for benchmark 1, for kinetic mixing, recovered using dynamic nested sampling under background scenarios $N_\mathrm{BG} = 1$ (left) and $N_\mathrm{BG} = 100$ (right), from one representative pseudo-data set. The five sampled parameters are the signal parameters $m_{A'}$ and $\epsilon$, and the nuisance parameters $\theta_S$, $\theta_B$, and $\sigma_B$ governing the signal and background systematic uncertainties. Dashed lines denote the posterior median with 68\% credible intervals; red points indicate the true injected values. Contours in the two-dimensional panels enclose the 68\% and 95\% credible regions.}
\label{fig:bayesian_cornerBM1}
\end{figure}
\begin{figure}
    \centering
\includegraphics[width=\linewidth]{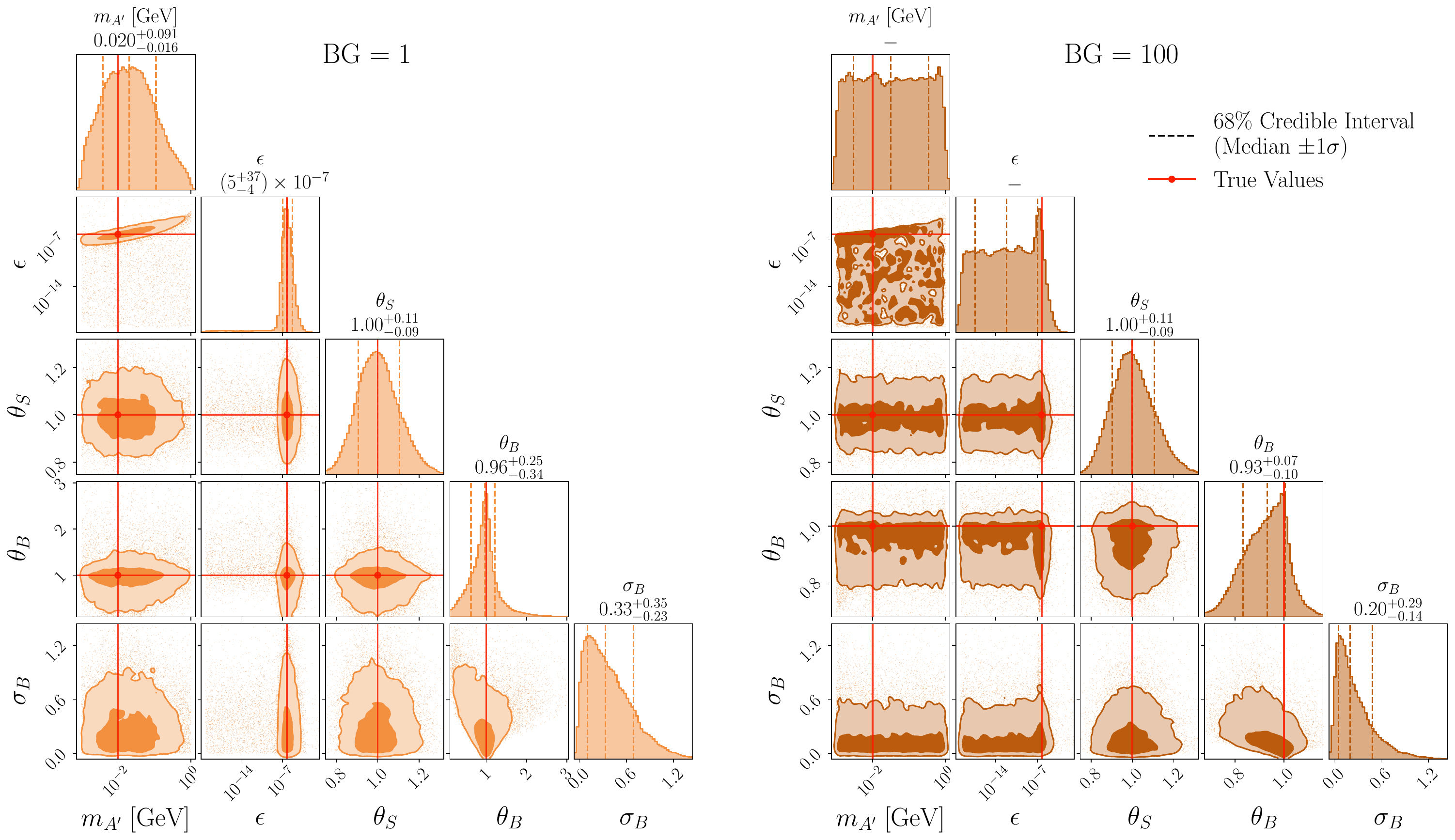}
\caption{The same as \figref{fig:bayesian_cornerBM1} but for benchmark 2. Dashes indicate that the posterior spans nearly the full prior range, rendering the obtained constraints prior dependent and thus not worth reporting.}
\label{fig:bayesian_cornerBM2}
\end{figure}
\begin{figure}
    \centering
\includegraphics[width=\linewidth]{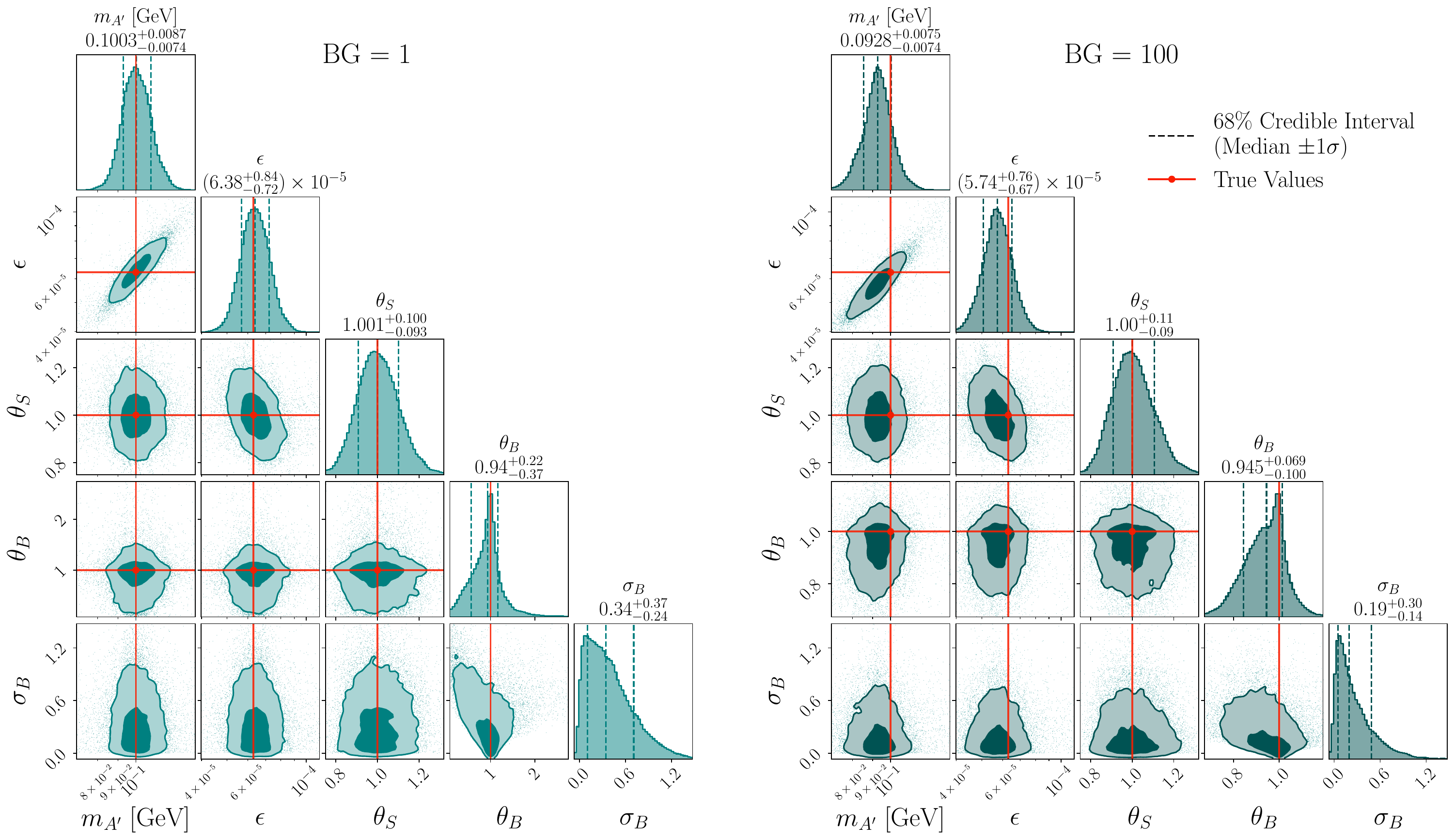}
\caption{The same as \figref{fig:bayesian_cornerBM1} but for benchmark 3.}
\label{fig:bayesian_cornerBM3}
\end{figure}
\begin{figure}
    \centering
\includegraphics[width=\linewidth]
{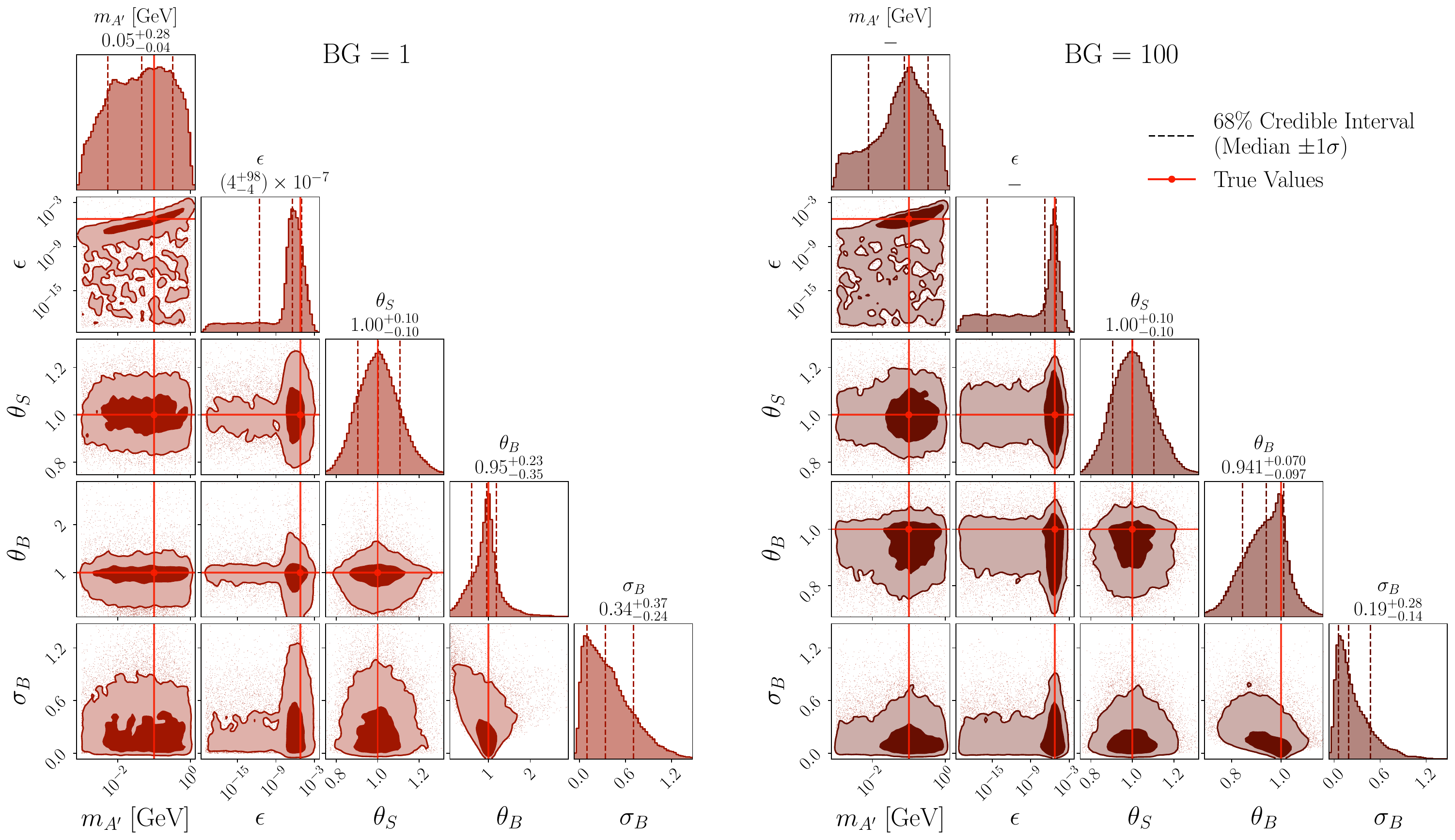}
\caption{The same as \figref{fig:bayesian_cornerBM1} but for benchmark 4. Dashes indicate that the posterior spans nearly the full prior range, rendering the obtained constraints prior dependent and thus not worth reporting.}
\label{fig:bayesian_cornerBM4}
\end{figure}
Figures~\ref{fig:bayesian_cornerBM1}--\ref{fig:bayesian_cornerBM4} show the marginalized posterior distributions for all five sampled parameters, the signal parameters $m_{A'}$ and $\epsilon$, and the nuisance parameters $\theta_S$, $\theta_B$, and $\sigma_B$, recovered from one representative pseudo-data set at each benchmark point under both background scenarios. The results divide naturally into two qualitatively distinct regimes: benchmarks for which the posterior is strongly informed by the data (benchmark 1 and 3), and benchmarks for which the signal parameters remain largely prior-dominated with too few signal events for strong inference (benchmark 2 and 4).

At benchmarks 1 and 3, the expected signal yield is sufficiently large that the likelihood retains strong dependence on both signal parameters. The recovered posteriors are well-localized and approximately Gaussian, with credible intervals of order $\sim10$\% of the reconstructed values on both $m_{A'}$ and $\epsilon$ in both background scenarios. The posterior medians are consistent with the injected values to within $1\sigma$, and the nuisance parameter $\theta_S$ is recovered with comparable precision across all cases ($\pm 10\%$) and remains largely uncorrelated with the signal parameters --- the signal yield is sufficiently high that the shape information in $m_{A'}$ and the overall normalization in $\epsilon$ are resolved independently of the signal uncertainty. The credible regions of $\theta_B$ shrink for the larger background scenario due to the increased number of background events.
The two-dimensional panels reveal a positive correlation between $m_{A'}$ and $\epsilon$: increasing $m_{A'}$ decreases the number of expected signal events, which can be compensated by a correlated increase in $\epsilon$ to maintain the same total event count. The $\sigma_B$ parameter, encoding the fractional uncertainty on the background rate, remains only weakly constrained relative to its prior in all scenarios, with the posterior peaking near zero and exhibiting a long upper 
tail. The data carries little direct information about the intrinsic scatter of the 
background in only one pseudo-data set.

At benchmarks 2 and 4, the coupling strength lies sufficiently far below the sensitivity threshold that the expected signal contribution is comparable to or smaller than the Poisson fluctuations in the background.
In this regime, the posterior distributions on $m_{A'}$ and $\epsilon$ are essentially uninformative: their posteriors span the entire prior range and the credible intervals are wide with respect to the prior range. The physical origin of these wide posteriors is that the signal rate is at such a small level, indistinguishable from background given Poisson fluctuations, such that the likelihood becomes independent of $\epsilon$ over this range. The data is therefore consistent with arbitrarily small couplings. Similarly, once $\epsilon$ is unconstrained, the spectral shape information that would otherwise localize $m_{A'}$ is lost, and the marginal posterior on $m_{A'}$ broadens to span the full prior range. 
This behaviour is more pronounced under $N_\mathrm{BG} = 100$, where the increased 
background fluctuations further obscure any residual signal. The background nuisance $\theta_B$ is, by contrast, well-recovered in the $N_\mathrm{BG} = 100$ scenario. 
Benchmark 2 under $N_\mathrm{BG} = 1$ occupies a borderline regime: the expected signal count of five events is sufficient to weakly inform the posterior, but insufficient to yield precise or symmetric credible intervals.

\subsection{Model Comparison}
\label{section:model_comparison}
\begin{figure}
    \centering
    \includegraphics[width=0.49\linewidth]{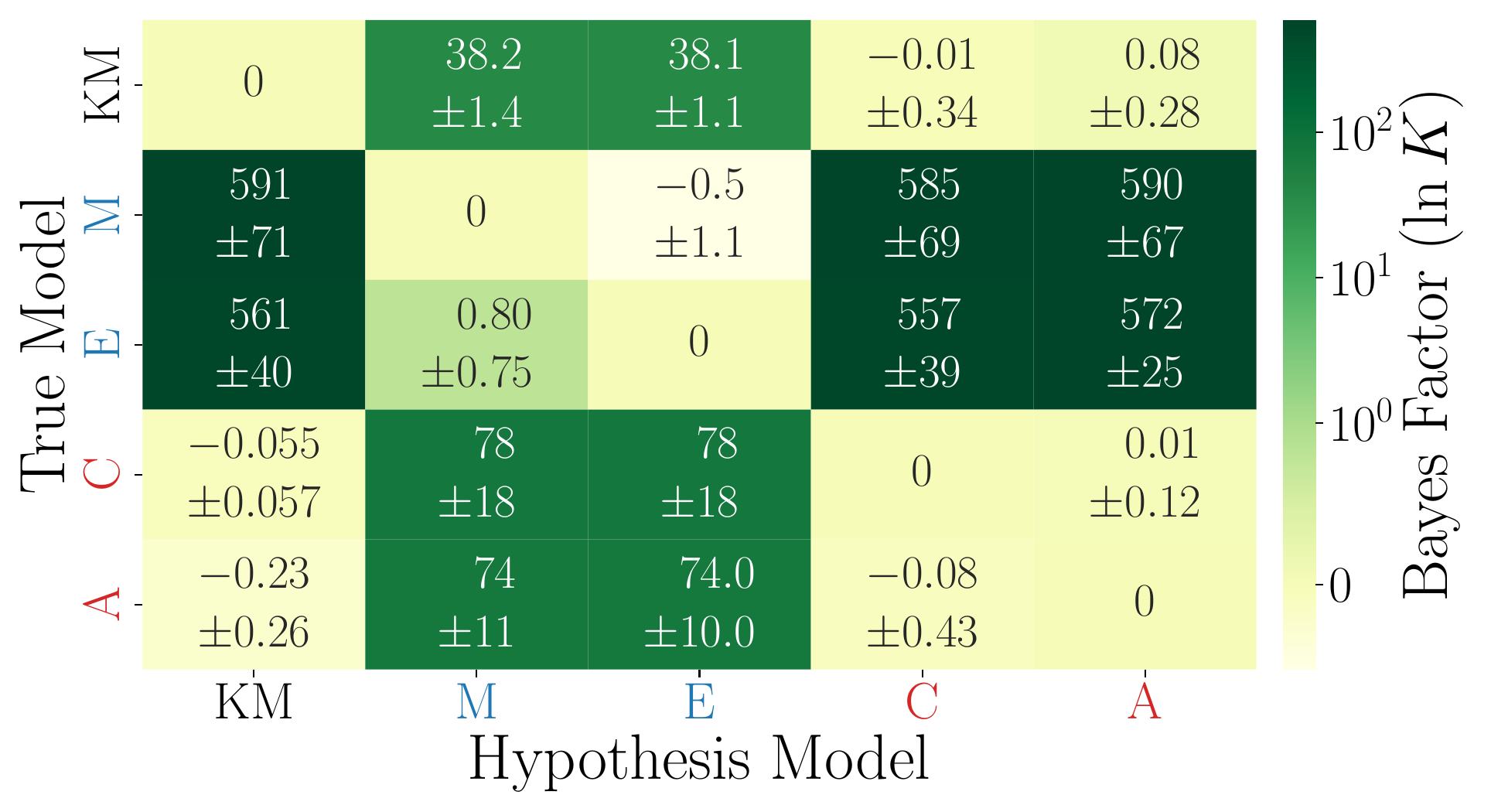}
    \includegraphics[width=0.49\linewidth]{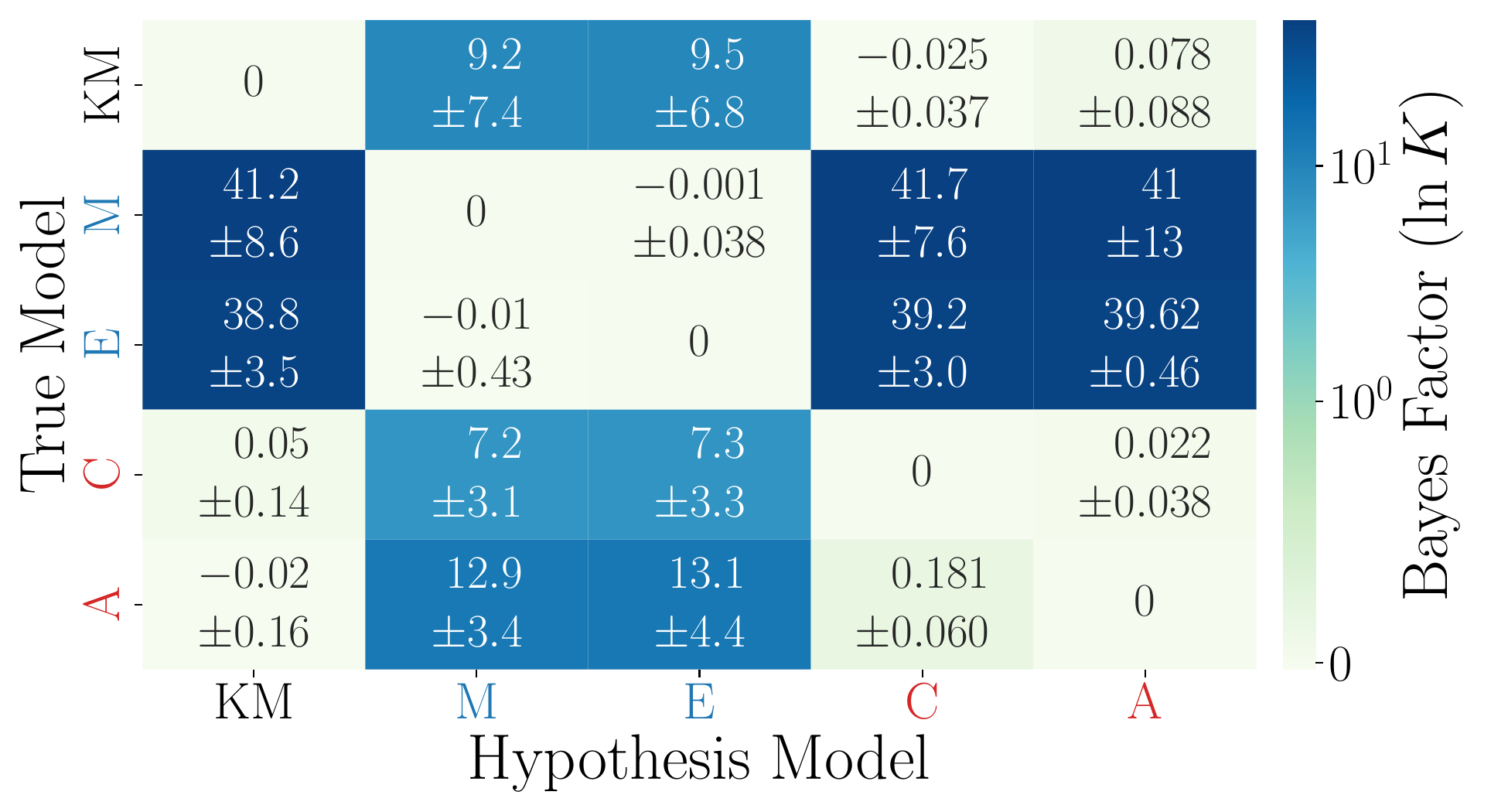}
    \caption{Bayes factor, $K$, for model discrimination using the full two-dimensional $(|p_T|_{e^-_f}, E_{e^-_f})$ recoil distribution (left) and the energy-only distribution (right). Each entry gives the median $\ln K$ over ten pseudo-data realizations, with the standard deviation of that ensemble shown beneath. Rows correspond to the true generative model from which pseudo-data are drawn; columns correspond to the hypothesis model whose evidence is evaluated relative to the generative model. Large positive values indicate decisive evidence against the hypothesis model, while small values indicate the two models are indistinguishable. Comparison of the two panels illustrates the gain in discriminating power afforded by incorporating the transverse momentum information alongside the energy. We take the benchmark 3 values ($m_{A'}=0.1$ GeV,$\mathcal{R}=2.5$) with $N_\text{BG}=1$ for the injected model parameters to generate the pseudo-data.}
    \label{fig:model_comparison}
\end{figure}
The Bayes factor heatmaps in Fig.~\ref{fig:model_comparison} quantify the ability of the LDMX likelihood-based framework to discriminate between the five signal models — kinetic mixing (KM), magnetic dipole (M), electric dipole (E), charge radius (C), and anapole moment (A) — as a function of which model generated the data. The left figure uses the full two-dimensional ($|p_T|_{e^-_f},E_{e^-_f}$) distributions, and the right figure uses only the energy distributions. Each cell reports the median $\log K$ over 10 pseudo-data realizations, with the ensemble standard deviation shown beneath. By construction, the diagonal entries vanish: when the hypothesis and generative models coincide, the Bayes factor is unity and $\log K = 0$. The couplings, $g_f$, used for this analysis are given in Table~\ref{tab:DEM_benchmarks}, which correspond to the couplings on the relic target for benchmark 3, along with the corresponding number of expected signal events. 
\renewcommand{\arraystretch}{1.2}
\begin{table}[h]
\centering
\caption{The coupling values, $g_f$, defined in \eqnref{eq:g_f}, and associated number of expected signal events $S$, on benchmark 3, $m_{A'} = 0.1$ GeV and $\mathcal{R} \equiv \frac{m_{\Ap}}{m_\text{DM}} = 2.5$, for all five dark photon models considered in this work.}
\label{tab:DEM_benchmarks}
\begin{tabular}{lccccc}
\toprule \toprule
& KM & M & E & C & A \\
\hline
$g_f$  & 5 $\times 10^{-6}$ & 8 $\times 10^{-4}$ & 8 $\times 10^{-4}$ & 0.06  & 0.06  \\
$S$ & 250 & 39000 & 39000 & 5300 &  5300  \\
\toprule \toprule
\end{tabular}
\end{table}

The M and E models predict identical recoil distributions and event yields, as do C and A; consequently, the corresponding off-diagonal pairs are mutually indistinguishable, with $\ln K \approx 0$ in both panels. The small fluctuations from 0 at these combinations are due to random fluctuations in the \madgraph Monte Carlo simulations. This degeneracy reflects the fact that the interaction Lagrangian for M and E differ only by a $\gamma^5$, and similarly for C and A, therefore the dark photon production cross sections are approximately equal in the relativistic limit where the electron mass can be neglected. This fact is also demonstrated by the similar values for the (M,E) and (C,A) row and column pairs on the chart when compared with other the other models.

The KM model shares the same recoil spectral shape as C and A but differs in overall signal normalization \cite{Catena:2025fsl}. Despite KM having approximately twenty times fewer 
signal events than C or A at the relic target, all three models remain nearly indistinguishable. This 
is because the free parameter $g_f$ in the hypothesis model can absorb the normalization difference during evidence evaluation, leaving only the spectral shape to discriminate. Below, we compute the Baye's factors with a relic density prior on $g_f$ to penalize fits away from the relic target. 

C and A models carry an additional momentum-suppression factor relative to M/E, which reduces their signal yields despite the larger values of $g_f$ required to reach the relic target. Due to these differences in the signal yield between models, notable asymmetry is visible in the off-diagonal discrimination power: generating pseudo-data from M or E and testing the KM hypothesis yields $\ln K \sim 600$ (two-dimensional) or $\ln K \sim 40$ (energy-only), whereas the reverse — generating from KM and testing M or E — gives $\ln K \sim 40$ (two-dimensional) or $\ln K \sim 10$ (energy-only). Since we are evaluating these models at the relic target benchmark, models with a higher signal yield at the relic target provide greater statistical power to discriminate against any competing hypotheses with sufficiently different signal shapes (such as C/A with M/E). This is because more event leads to better shape reconstruction and thus better discriminating power between signals of inherently different distribution shapes.
The larger ensemble variance in the M and E rows of the two-dimensional panel reflects the sensitivity of the Bayes factor to Monte Carlo fluctuations at higher signal yields.

Comparing the two panels, the two-dimensional $(|p_T|_{e^-_f}, E_{e^-_f})$ analysis affords substantially greater discriminating power than the energy-only analysis, approximately by an order of magnitude in $\log K$. This gain demonstrates the value of retaining the full two-dimensional recoil information and motivates the use of the joint $(|p_T|_{e^-_f}, E_{e^-_f})$ likelihood. It is also interesting to note that the $|p_T|_{e^-_f}$-only equivalent analysis delivers less discriminating power, with a factor of $\sim 2-3$ less in $\log K$ compared with the energy-only analysis.

\begin{figure}
    \centering
    \includegraphics[width=0.5\linewidth]{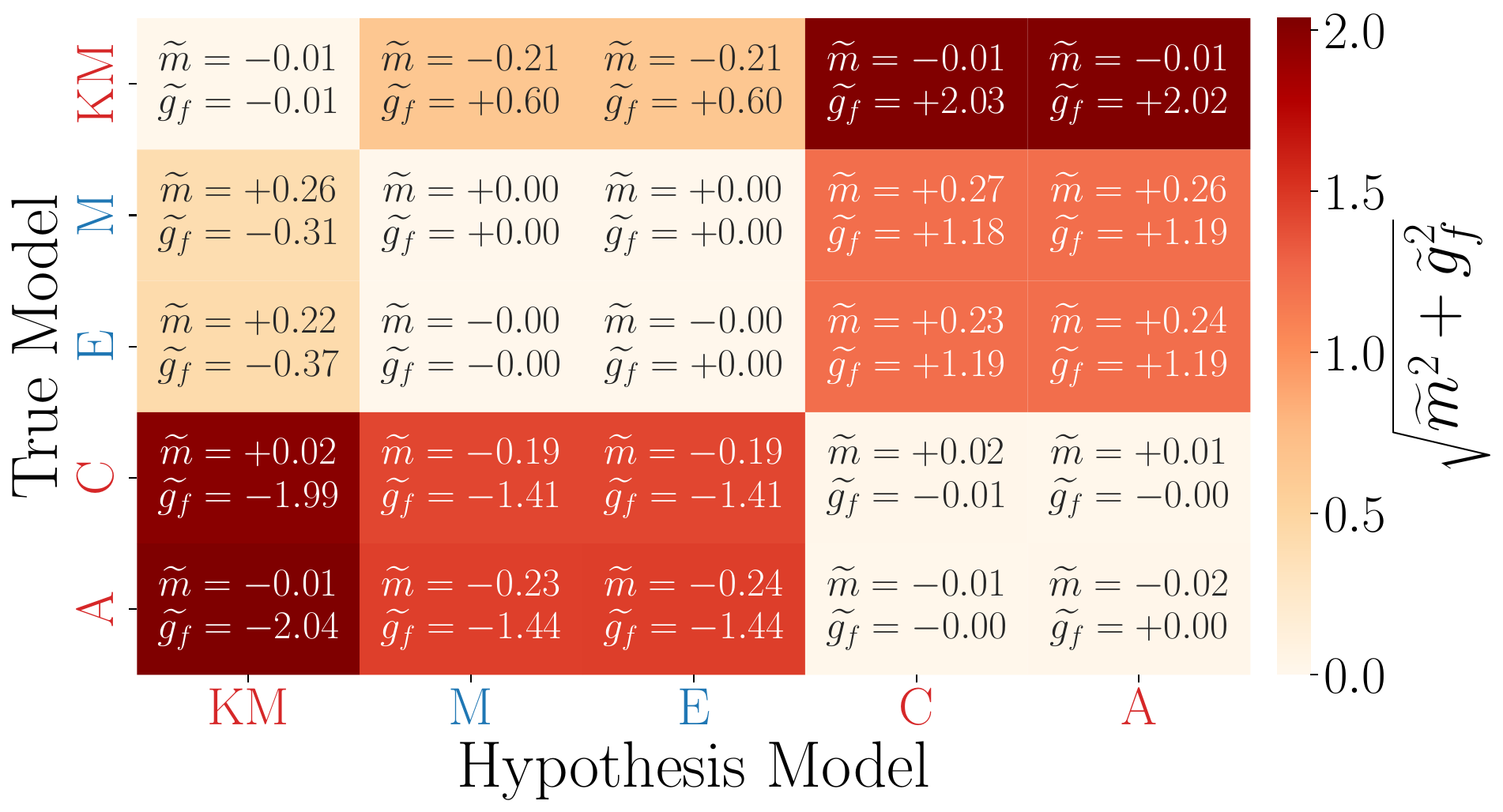}
    \caption{Grid of biases of the inferred model parameters $m_{A'}$ and $g_f$ as defined in \eqnref{eq:bias2}, assuming the hypothesis model on the x-axis and the true injected model on the y-axis. The colour-bar represents a combined bias across both model parameters, where a darker colour indicates less accuracy. Biases are the medians across 10 pseudo-data sets. We take the benchmark 3 values ($m_{A'}=0.1$ GeV,$\mathcal{R}=2.5$) for the injected model parameters to generate the pseudo-data, where $g_f$ are the values on the relic targets.}
    \label{fig:misspecification}
\end{figure}
Figure~\ref{fig:misspecification} reports the parameter reconstruction bias when 
pseudo-data generated under the true model (rows) are fitted with a miss-specified hypothesis model (columns). The bias of the predicted parameters, showing the accuracy of the inferred model parameters, are reported as,
\begin{equation}
\begin{aligned}
    &\tilde{m}\equiv \log_{10}\Big( \hat{m}_{A'}/\text{GeV} \Big) - \log_{10}\Big(m_{A',\text{true}}/\text{GeV}\Big) \\
    &\tilde{g}_f \equiv \log_{10}\hat{g}_f - \log_{10}g_{f,\text{true}},
    \label{eq:bias2}
\end{aligned}
\end{equation}
where under this definition, $\tilde{g}_f = -2$ means that the sampler has fit a coupling strength two orders of magnitude lower than the true value $g_{f,\text{true}}$. We take the quantity $\sqrt{\tilde{m}^2+\tilde{g}_f^2}$ to represent the overall reconstruction bias. 

The colour of each cell encodes the combined bias 
$\sqrt{\tilde{m}^2 + \tilde{g}_f^2}$, with the individual biases $\tilde{m}$ and 
$\tilde{g}_f$ printed explicitly. Several distinct regimes are visible.

Along the diagonal, both biases are consistent with zero, confirming that the sampler 
accurately recovers the injected parameters when the hypothesis model is correctly 
specified. The same holds for the M/E and C/A off-diagonal pairs: since these models 
predict identical recoil distributions and event yields, fitting with the wrong member 
of each pair incurs negligible bias. This is consistent with the near-zero Bayes factors 
for these combinations in Fig.~\ref{fig:model_comparison}.

The most striking feature of the miss-specification matrix is the large coupling bias that 
arises when KM data are fitted with C or A (and vice versa). When the true model is KM 
and the hypothesis is C or A, the inferred coupling is approximately two orders of 
magnitude above the true KM value, while the mass bias 
remains small ($\tilde{m} \approx 0$). The converse — fitting C or A data with the KM 
hypothesis — yields $\tilde{g}_f \approx -2.0$ with similarly negligible mass bias. 
This pattern reflects the shared spectral shape between KM and C/A: because the recoil distributions are degenerate in shape, the sampler correctly 
recovers the mediator mass but cannot determine the overall normalization without 
knowledge of the true coupling structure. Crucially, these large coupling biases occur 
precisely in the regime where the Bayes factor is near zero — the models are 
statistically indistinguishable, yet the inferred parameters are grossly incorrect. 
This illustrates that a near-zero Bayes factor does not imply accurate parameter 
recovery when two models are shape-degenerate but differ in normalization.

When M or E data are fitted with KM (or vice versa), moderate biases appear in both parameters, corresponding to a mild overestimate of the mediator mass and an underestimate of the 
coupling. When M or E data are fitted with C or A (or vice 
versa), not only is there a mild mass bias but the coupling bias is large, reflecting the 
order-of-magnitude difference in signal yield between the models at the relic target.

To further discriminate between these competing signal hypotheses, especially ones with the same shape, we implement the relic density prior on $g_f$ as described in \secref{sec:model_comparison} which penalizes fits that are not compatible with the freeze-out relic target curve.
\begin{figure}
    \centering
    \includegraphics[width=0.5\linewidth]{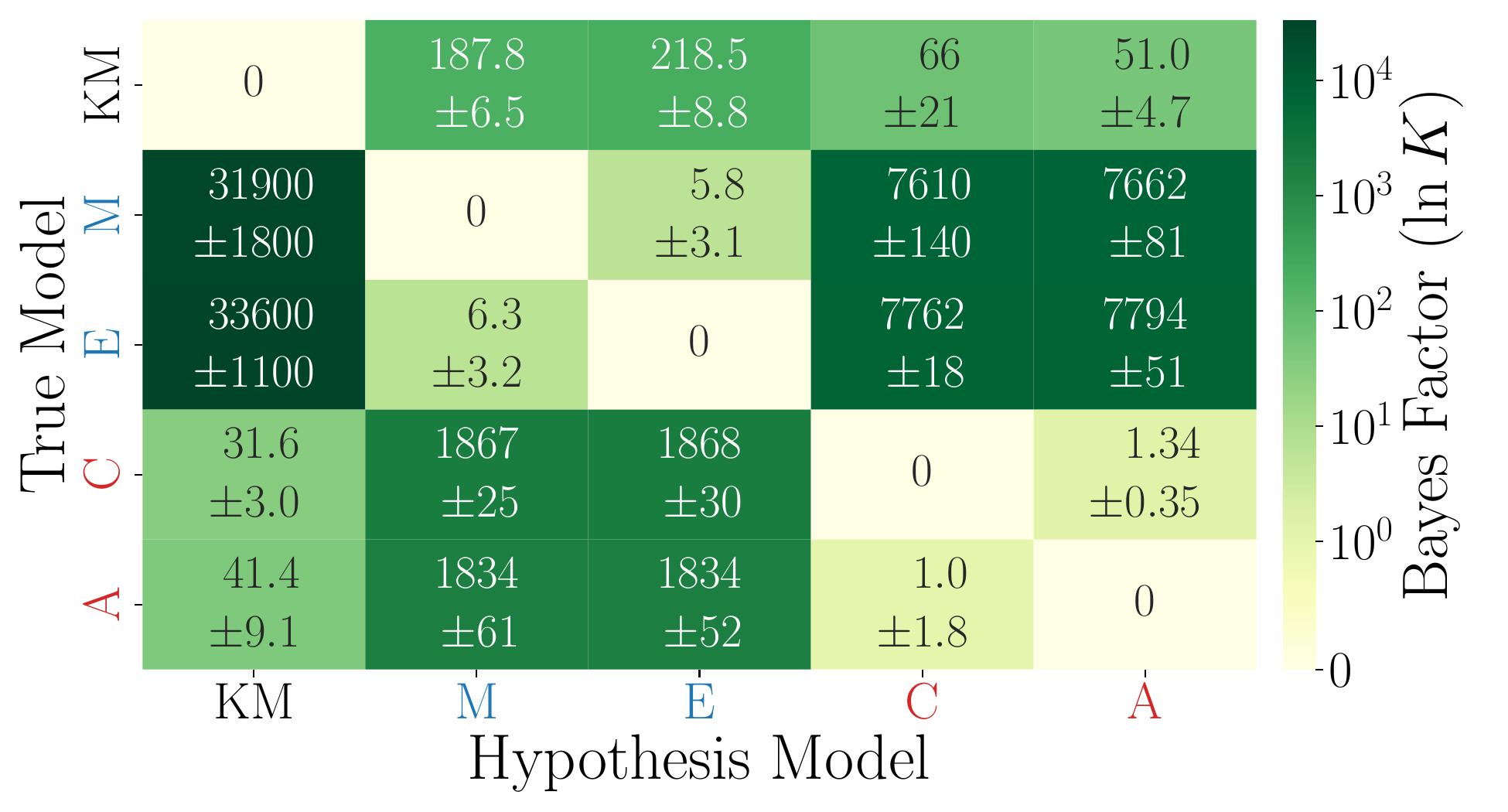}
    \caption{Same as \figref{fig:model_comparison} but with the relic density prior on $g_f$. Model parameters are fixed to $m_{A'}=0.1$ GeV, $\mathcal{R}=3$, and $g_f$ is taken to be the values on the relic target from Fig. 1 of \cite{Catena:2025fsl}.}
    \label{fig:model_comparison_relicpior}
\end{figure}
The previously indistinguishable model combination, KM and C/A, now have large Bayes factors, indicating that the relic target prior constraint causes them to be statistically differentiable. Since the total number of signal events predicted at LDMX for each model on the $\mathcal{R}=3$ relic target is larger than for $\mathcal{R}=2.5$, the Bayes factors are consistently larger for all cells. Therefore, it is the statistical variation in the \madgraph generated signal that is causing the Bayes factors to be of order $1$ for the pairs (M,E) and (C,A), and we should not take this to indicate they are distinguishable.

\section{Conclusion}
\label{section:conclusion}

We present the evaluation of the exclusion and discovery potential, and the parameter estimation and model selection power of the LDMX missing momentum experiment~\cite{LDMX:2025bog} for light complex scalar DM, across two representative background scenarios, $N_\mathrm{BG} = 1$ and $N_\mathrm{BG} = 100$ background events. We evaluate the discovery potential and parameter inference capability on a set of four benchmark values of $(m_{A'},\epsilon)$, the dark photon mediator mass and the kinetic mixing parameter, on the relic target curve, for $\mathcal{R}  \equiv \frac{m_{A'}}{m_\text{DM}} = 2.5$ and $2.2$. In previous literature such as \cite{Berlin:2018bsc}, $\mathcal{R}$ is routinely taken to be 3, however recent electron recoil direct detection experiments exclude this mass ratio \cite{Krnjaic:2025noj}, motivating our choice of benchmarks.  Model selection is evaluated on one of the benchmarks.
Our analysis projects that, in the case of a null result, the relic target is excluded for $\mathcal{R} \gtrsim 2.2$ across the mass range $m_{A'} \in [10^{-3},\, 0.2]$~GeV, with the precise reach depending on $\mathcal{R}$ and the background level. 
The discovery significance is evaluated at each benchmark point, revealing that the $\mathcal{R}=2.5$ benchmarks are \textit{discoverable}, while the $\mathcal{R}=2.2$ benchmarks are just out of $5\sigma$ reach for low background.

Parameter estimation of $\epsilon$ and $m_{A'}$ is performed using both frequentist maximum likelihood and Bayesian posterior inference. We find that LDMX is capable of precise and accurate parameter estimation along the $\mathcal{R}=2.5$ relic target: benchmarks~1 and~3 are recovered within $1\sigma$ of the injected values, with relative uncertainties of order $10\%$, across both background scenarios. Benchmarks~2 and~4 exhibit reduced accuracy and larger uncertainties, and in some cases yield uninformative constraints for the $N_\mathrm{BG} = 100$ scenario. Relative biases and half-interval widths quantify accuracy and precision, respectively, in the frequentist treatment. Meanwhile, the Bayesian posteriors provide a natural quantification of the uncertainties on the parameters and the expected bias, without relying on the asymptotic approximations required for frequentist interval construction. Overall, the main conclusions from the frequentist and Bayesian analyses are similar to each other.

Extending beyond kinetic mixing, we consider additional dark photon models coupling to SM fermions through higher-order electromagnetic moments~\cite{Catena:2025fsl} -- DEM models. In particular, we study the magnetic (M) and electric (E) dipole moments, the charge radius (C), and the anapole moment (A), and compute Bayes factors for model discrimination between these signal hypotheses. At benchmark 3, the two-dimensional $(|p_T|_{e^-_f}, E_{e^-_f})$ analysis yields Bayes factors of order $10$--$100$ between distinguishable model pairs, approximately an order of magnitude larger than those obtained from an energy-only analysis; in particular, KM, C, and A are each distinguishable from E and M, and vice versa. We further evaluate the bias incurred when fitting data generated under one model to a different, incorrectly assumed model, quantifying the extent to which model misspecification propagates into biased parameter inference. Finally, incorporating a relic density prior increases the Bayes factor across all model pairs, rendering some previously indistinguishable signal hypotheses statistically distinguishable.

Taken together, these results demonstrate that LDMX Phase II will not only probe the thermal relic target for light complex scalar DM not yet excluded by recent direct detection results~\cite{DAMIC-M:2025luv,PandaX:2025rrz} with strong discovery and exclusion sensitivity, but, in the event of an excess, will be capable of directly characterizing the underlying dark sector interaction. The statistical framework developed here exploits the two-dimensional recoil electron distributions within a combined frequentist and Bayesian likelihood treatment, incorporating signal and background modelling with their associated systematic uncertainties. On the frequentist side, we demonstrated that the asymptotic $\tfrac{1}{2}\chi^2$ approximation to the distribution of the profile likelihood ratio test statistic breaks down due to the low counts per bin, and that full generation of pseudo-experiments are essential for reliable exclusion and discovery statements. The statistical likelihood-based framework is directly applicable to other missing momentum experiments, such as Lohengrin~\cite{Bechtle:2024atq} and DarkSHINE~\cite{DarkSHINE:2024guq}, in addition to other DM models in which a signal is expected, and is designed for direct application to real LDMX data once available. It can furthermore be naturally extended to a combined likelihood incorporating results from other probes of light DM and dark photons, enabling joint constraints across the broader light DM experimental program. As LDMX moves toward construction and commissioning, and as the viable thermal relic parameter space is progressively narrowed by direct detection experiments, the tools developed here will be essential for translating raw event counts into statistically robust physics conclusions.

\acknowledgments
We thank the LDMX group at Lund University for many valuable discussions throughout the development of this work.
R.C. and T.G. have been funded by the Knut and Alice Wallenberg Foundation, and performed their research within the ``Light Dark Matter'' project (Dnr. KAW 2019.0080).

\appendix

\section{Detector Resolution}
\label{section:smear}

To account for the finite detector resolution, the predicted signal and background distributions in the recoil electron energy and transverse momentum magnitude are smeared. This procedure models the migration of events between bins due to measurement uncertainties.
The detector resolutions in $E_{e^-_f}$ and $|p_T|_{e^-_f}$ are treated as independent and are modelled using Gaussian response functions.

The $|p_T|_{e^-_f}$ resolution, $\sigma_{p_T}(p_T)$, is obtained from values provided in Fig. 3.26 of \cite{LDMX:2025bog}. These values are interpolated linearly to provide a continuous function.
The energy resolution is parameterized as,
\begin{equation}
    \sigma_E(E) = \sqrt{s^2 E + c^2 E^2 + n^2},
\end{equation}
where $s$, $c$, and $n$ are the stochastic, constant, and noise terms, respectively. In this analysis, we take the values $s = 0.2\: \text{GeV}^{1/2}$, $c = 0.03$, $n = 0 \: \text{GeV}$, motivated by the values reported in Table 3.10 of \cite{LDMX:2025bog}.

The smearing is implemented as a bin-to-bin migration using a response matrix. For each observable, we compute the probability that an event originating in a given true bin $j$ is reconstructed in bin $i$.
For a set of bin edges $x_i$, the probability of migrating from a bin centred at $x_j$ into bin $i$ is given by integrating the Gaussian response over the bin,
\begin{equation}
P_{ij} = \Phi(x_{i+1}; x_j, \sigma(x_j)) - \Phi(x_i; x_j, \sigma(x_j)),
\end{equation}
where $\Phi$ is the cumulative distribution function of a normal distribution with mean $x_j$ and width $\sigma(x_j)$.
This procedure is applied independently to $E_{e^-_f}$ and $|p_T|_{e^-_f}$, yielding two response matrices:
$P^{E}$ and $P^{p_T}$.
Assuming factorization of the detector response, the full two-dimensional response matrix is constructed as the Kronecker product $R = P^{E} \otimes P^{p_T}$. This method approximately preserves normalization, up to edge effects where events may migrate beyond the boundaries of the defined bin range.

\begin{figure}
    \centering
    \includegraphics[width=\linewidth]{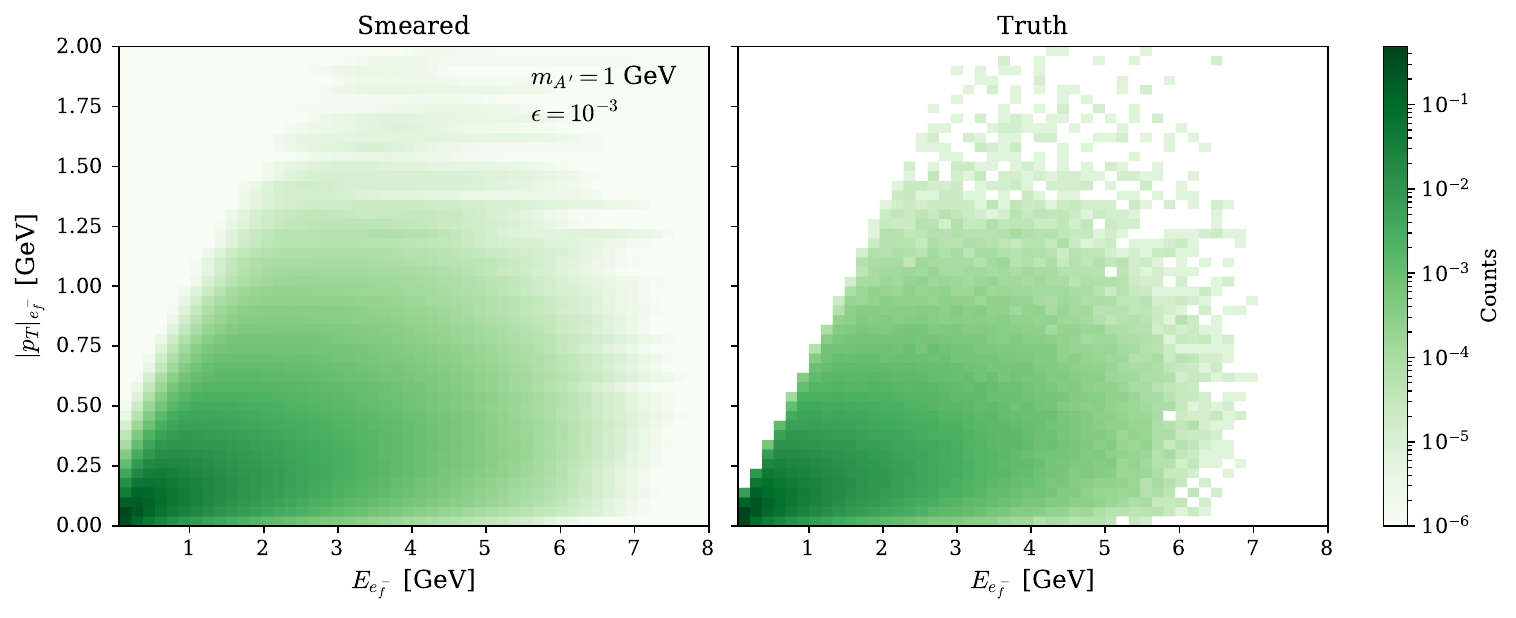}
    \caption{The smeared (left) and truth (right) two-dimensional signal distributions in the $|p_T|_{e^-_f}$ vs $E_{e^-_f}$ plane at a representative $m_{A'}$ and $\epsilon$.}
    \label{fig:smear}
\end{figure}

A comparison between the unsmeared “truth” and smeared signal distributions is shown in \figref{fig:smear}. The effect of the detector resolution is to broaden the distributions and wash out sharp features. This results in a migration of events toward neighbouring bins and a general smoothing of the spectrum.
This smearing procedure is applied consistently to both signal and background predictions before performing statistical analysis.

\section{Test Statistic Distributions}
\label{section:test_statistic_dists}

\begin{figure}
    \centering
    \includegraphics[width=0.74\linewidth]{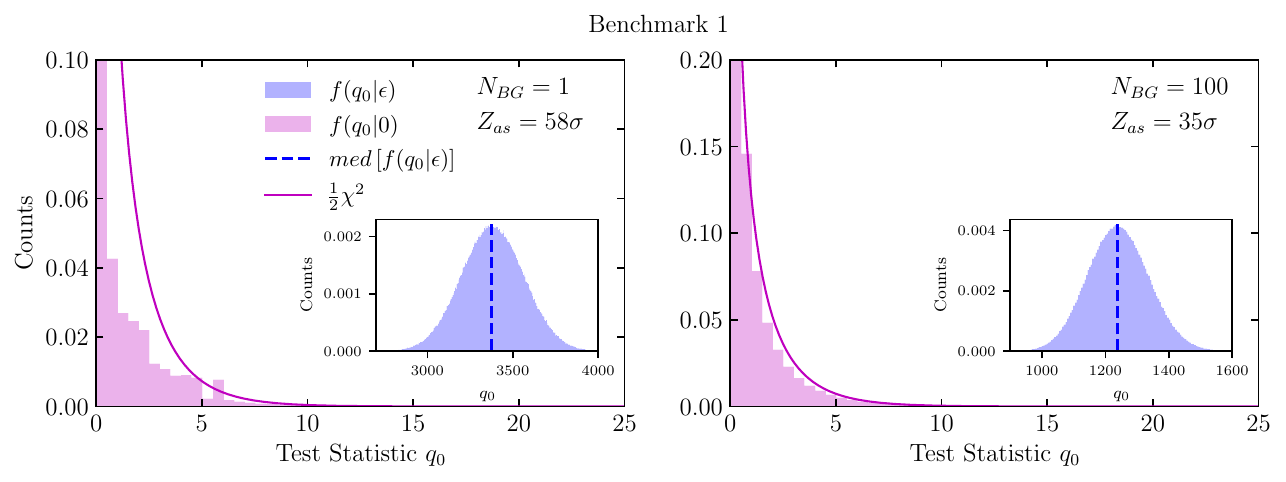}
        \includegraphics[width=0.74\linewidth]{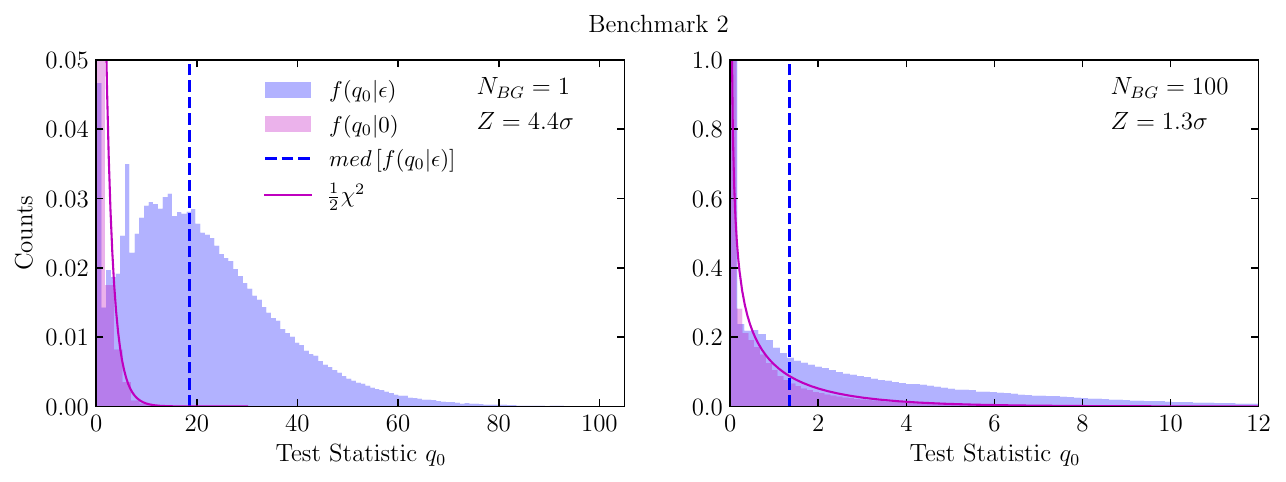}
    \includegraphics[width=0.74\linewidth]{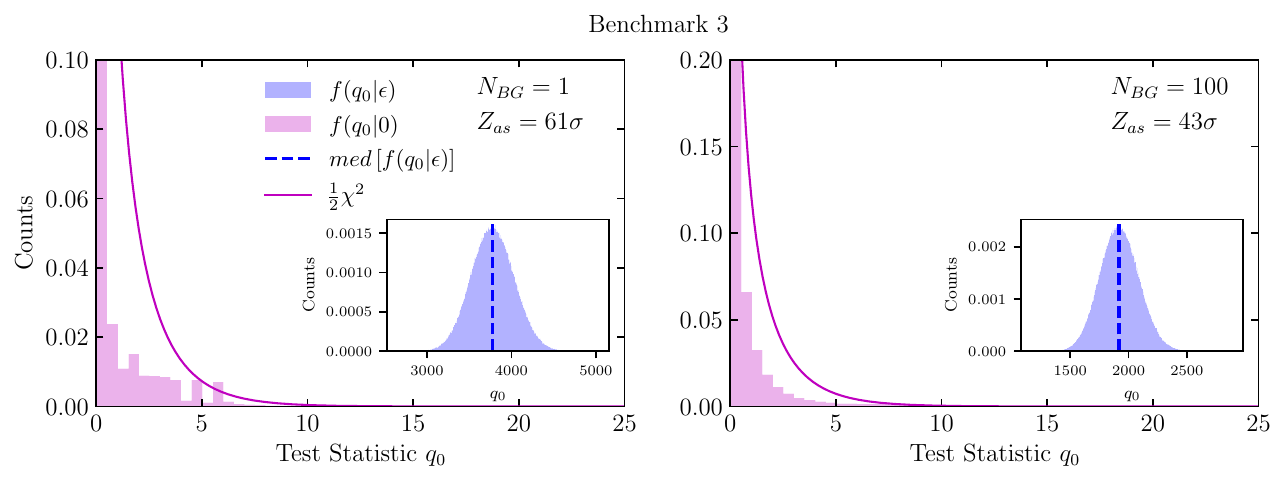}
    \includegraphics[width=0.74\linewidth]{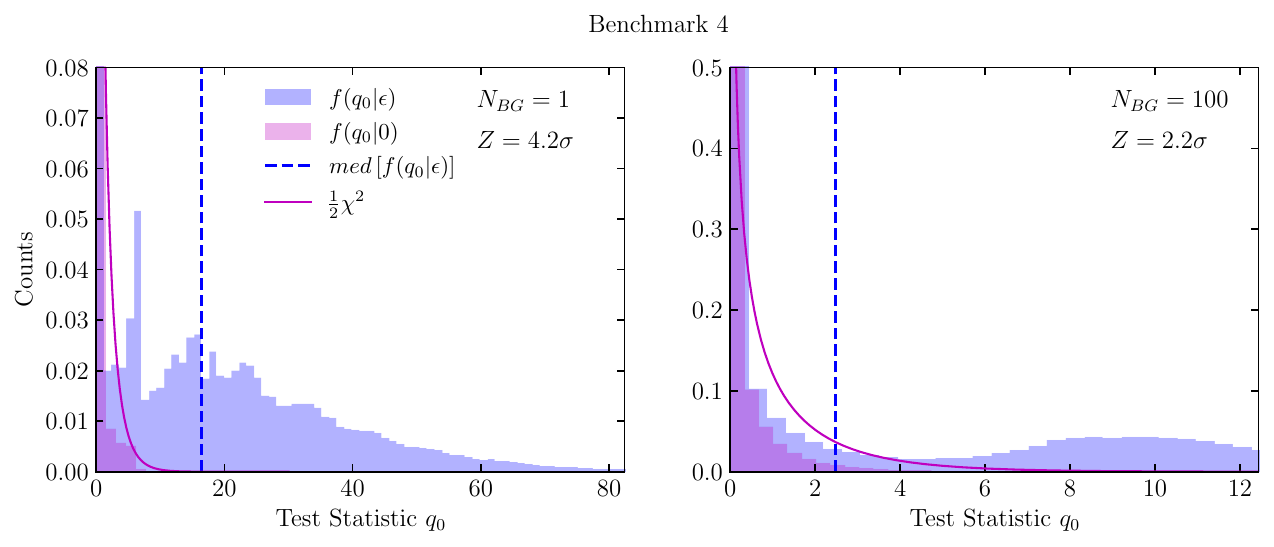}
    \caption{The distributions of the test statistic $q_0$ for discovery for each of the four benchmark points (rows), evaluated for background scenarios $N_\text{BG}=1$ (left) and $N_\text{BG}=100$ (right). The distribution of $q_0$ under background only, $f(q_0|0)$, is plotted in pink, while the distribution under signal-plus-background, $f(q_0|\epsilon)$, is in blue. The asymptotic half $\chi^2$ approximation of $f(q_0|0)$ derived in \cite{Likelihoods} is drawn in magenta. The blue dashed vertical line is the median of $f(q_0|\epsilon)$, used in the computation of the median discovery significance, $Z$, or the asymptotic $Z_\text{as}$ for $Z\gg 5$.
    $\sim10^8$ pseudo-experiments were generated to produce the distributions under background only and $\sim 10^6$ for background-plus-signal.}
    \label{fig:discovery_q}
\end{figure}

\begin{figure}
    \centering
    \includegraphics[width=0.74\linewidth]{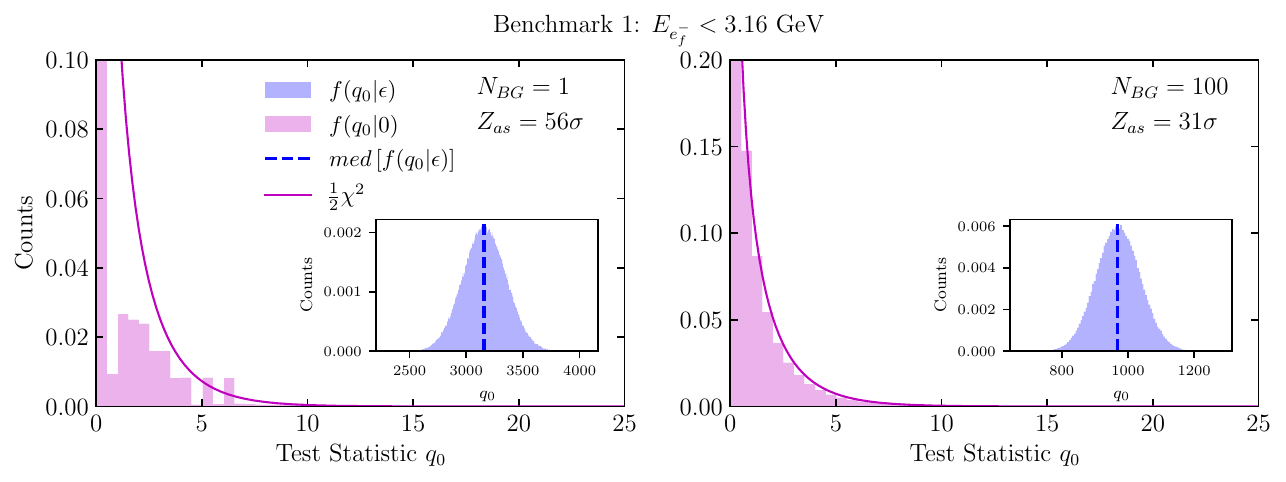}
        \includegraphics[width=0.74\linewidth]{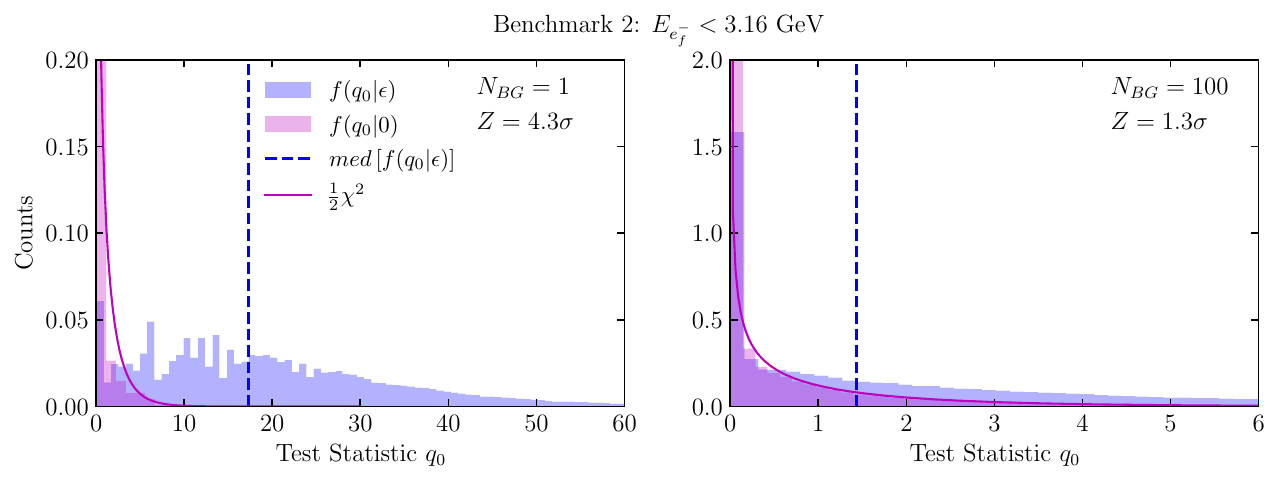}
    \includegraphics[width=0.74\linewidth]{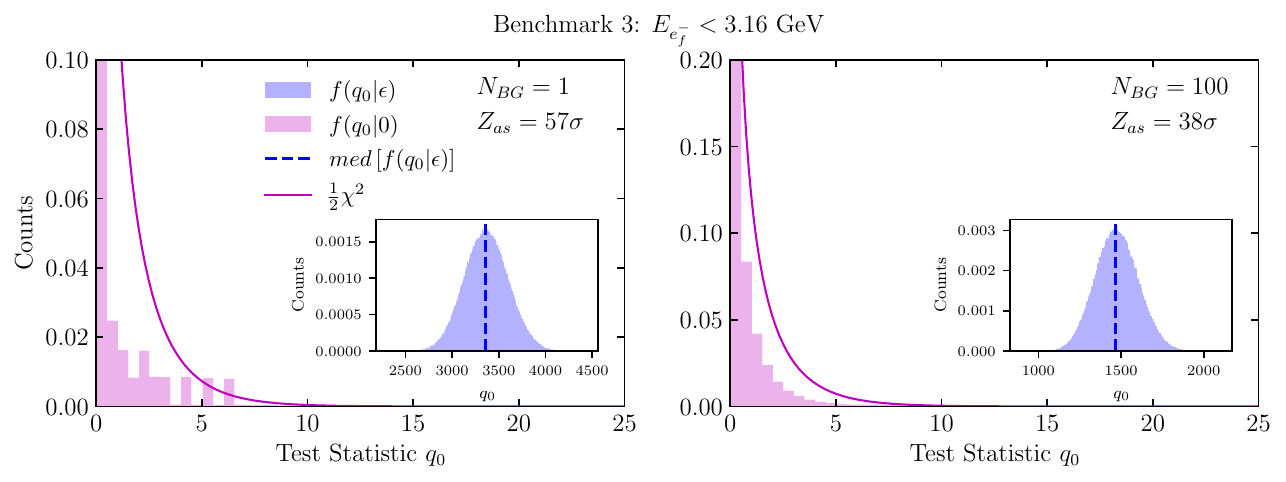}
    \includegraphics[width=0.74\linewidth]{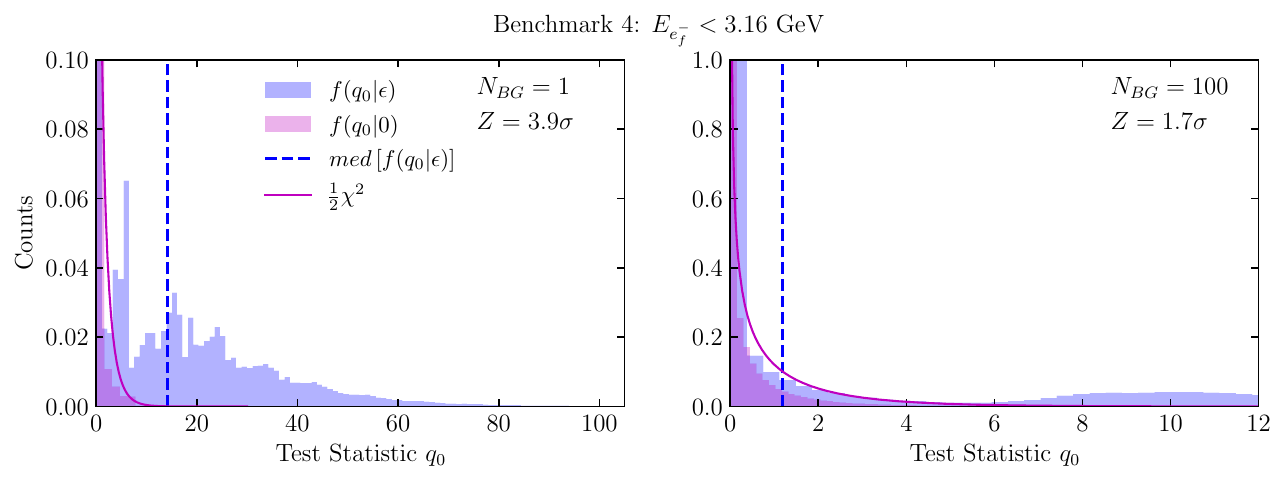}
    \caption{The same as \figref{fig:discovery_q}, but with the energy cut $E_{e^-_f}<3.16$ GeV applied.}
    \label{fig:discovery_q_Ecut}
\end{figure}

Figs.~\ref{fig:discovery_q} and \ref{fig:discovery_q_Ecut} show the distributions of the test statistic $q_0$ for each benchmark point and both background scenarios, with and without the energy cut, respectively. 
The toy distributions confirm that the asymptotic $\tfrac{1}{2}\chi^2$ approximation does not accurately describe $f(q_0|0)$ in this low-count regime, validating the use of pseudo-experiments for this analysis.

\clearpage
\bibliography{biblio}

@article{Likelihoods,
	doi = {10.1140/epjc/s10052-011-1554-0},
  
	url = {https://doi.org/10.1140%2Fepjc%2Fs10052-011-1554-0},
  
	year = 2011,
	month = {feb},
  
	publisher = {Springer Science and Business Media {LLC}
},
  
	volume = {71},
  
	number = {2},
  
	author = {Glen Cowan and Kyle Cranmer and Eilam Gross and Ofer Vitells},
  
	title = {Asymptotic formulae for likelihood-based tests of new physics},
  
	journal = {The European Physical Journal C}
}

@article{DMsignals_MM,
	doi = {10.1103/physrevd.103.035030},
  
	url = {https://doi.org/10.1103%2Fphysrevd.103.035030},
  
	year = 2021,
	month = {feb},
  
	publisher = {American Physical Society ({APS})},
  
	volume = {103},
  
	number = {3},
  
	author = {Nikita Blinov and Gordan Krnjaic and Douglas Tuckler},
  
	title = {Characterizing dark matter signals with missing momentum experiments},
  
	journal = {Physical Review D}
}

@article{Izaguirre_2015,
	doi = {10.1103/physrevd.91.094026},
  
	url = {https://doi.org/10.1103%2Fphysrevd.91.094026},
  
	year = 2015,
	month = {may},
  
	publisher = {American Physical Society ({APS})},
  
	volume = {91},
  
	number = {9},
  
	author = {Eder Izaguirre and Gordan Krnjaic and Philip Schuster and Natalia Toro},
  
	title = {Testing {GeV}-scale dark matter with fixed-target missing momentum experiments},
  
	journal = {Physical Review D}
}

@article{Conrad:2002kn,
    author = "Conrad, Jan and Botner, O. and Hallgren, A. and Perez de los Heros, Carlos",
    title = "{Including systematic uncertainties in confidence interval construction for Poisson statistics}",
    eprint = "hep-ex/0202013",
    archivePrefix = "arXiv",
    doi = "10.1103/PhysRevD.67.012002",
    journal = "Phys. Rev. D",
    volume = "67",
    pages = "012002",
    year = "2003"
}

@book{Jeffreys:1939xee,
    author = "Jeffreys, Harold",
    title = "{The Theory of Probability}",
    isbn = "978-0-19-850368-2, 978-0-19-853193-7",
    series = "Oxford Classic Texts in the Physical Sciences",
    year = "1939"
}

@article{Bechtle:2024atq,
    author = "Bechtle, Philip and others",
    title = "{A proposal for the Lohengrin experiment to search for dark sector particles at the ELSA Accelerator}",
    eprint = "2410.10956",
    archivePrefix = "arXiv",
    primaryClass = "hep-ex",
    doi = "10.1140/epjc/s10052-025-14257-z",
    journal = "Eur. Phys. J. C",
    volume = "85",
    number = "5",
    pages = "600",
    year = "2025"
}

@inproceedings{Lista:2016chp,
    author = "Lista, Luca",
    title = "{Practical Statistics for Particle Physicists}",
    booktitle = "{2016 European School of High-Energy Physics}",
    eprint = "1609.04150",
    archivePrefix = "arXiv",
    primaryClass = "physics.data-an",
    doi = "10.23730/CYRSP-2017-005.213",
    pages = "213--258",
    year = "2017"
}

@article{LDMX:2025ixw,
    author = "{\r{A}}kesson, Torsten and others",
    collaboration = "LDMX",
    title = "{Sensitivity of an early dark matter search using the electromagnetic calorimeter as a target for the Light Dark Matter eXperiment}",
    eprint = "2508.08121",
    archivePrefix = "arXiv",
    primaryClass = "hep-ex",
    reportNumber = "FERMILAB-PUB-25-0583-CSAID-PPD",
    doi = "10.1007/JHEP12(2025)150",
    journal = "JHEP",
    volume = "12",
    pages = "150",
    year = "2025"
}

@article{LDMX:2018cma,
    author = "{\r{A}}kesson, Torsten and others",
    collaboration = "LDMX",
    title = "{Light Dark Matter eXperiment (LDMX)}",
    eprint = "1808.05219",
    archivePrefix = "arXiv",
    primaryClass = "hep-ex",
    reportNumber = "FERMILAB-PUB-18-324-A, SLAC-PUB-17303",
    month = "8",
    year = "2018"
}

@article{LDMX:2025bog,
    author = "Akesson, Torsten and others",
    collaboration = "LDMX",
    title = "{LDMX - The Light Dark Matter eXperiment}",
    eprint = "2508.11833",
    archivePrefix = "arXiv",
    primaryClass = "hep-ex",
    reportNumber = "FERMILAB-PUB-25-0605-CSAID-PPD",
    month = "8",
    year = "2025"
}

@article{Berlin:2018bsc,
    author = "Berlin, Asher and Blinov, Nikita and Krnjaic, Gordan and Schuster, Philip and Toro, Natalia",
    title = "{Dark Matter, Millicharges, Axion and Scalar Particles, Gauge Bosons, and Other New Physics with LDMX}",
    eprint = "1807.01730",
    archivePrefix = "arXiv",
    primaryClass = "hep-ph",
    reportNumber = "FERMILAB-PUB-18-310-A, SLAC-PUB-17297",
    doi = "10.1103/PhysRevD.99.075001",
    journal = "Phys. Rev. D",
    volume = "99",
    number = "7",
    pages = "075001",
    year = "2019"
}

@article{Fabbrichesi:2020wbt,
    author = "Fabbrichesi, Marco and Gabrielli, Emidio and Lanfranchi, Gaia",
    title = "{The Dark Photon}",
    eprint = "2005.01515",
    archivePrefix = "arXiv",
    primaryClass = "hep-ph",
    doi = "10.1007/978-3-030-62519-1",
    month = "5",
    year = "2020"
}

@article{Stueckelberg:1938hvi,
    author = "Stueckelberg, E. C. G.",
    title = "{Interaction energy in electrodynamics and in the field theory of nuclear forces}",
    doi = "10.5169/seals-110852",
    journal = "Helv. Phys. Acta",
    volume = "11",
    pages = "225--244",
    year = "1938"
}

@article{Holdom:1985ag,
    author = "Holdom, Bob",
    title = "{Two U(1)'s and Epsilon Charge Shifts}",
    reportNumber = "UTPT-85-30",
    doi = "10.1016/0370-2693(86)91377-8",
    journal = "Phys. Lett. B",
    volume = "166",
    pages = "196--198",
    year = "1986"
}

@incollection{Berger:2001zbn,
  author    = {Berger, James O. and Pericchi, Luis R.},
  title     = {Objective {Bayesian} Methods for Model Selection: Introduction and Comparison},
  booktitle = {Model Selection},
  editor    = {Lahiri, P.},
  series    = {IMS Lecture Notes--Monograph Series},
  volume    = {38},
  pages     = {135--207},
  publisher = {Institute of Mathematical Statistics},
  address   = {Beachwood, OH},
  year      = {2001},
  doi       = {10.1214/lnms/1215540968}
}

@article{Higson:2018cwj,
    author = "Higson, Edward and Handley, Will and Hobson, Michael and Lasenby, Anthony",
    title = "{Dynamic nested sampling: an improved algorithm for parameter estimation and evidence calculation}",
    eprint = "1704.03459",
    archivePrefix = "arXiv",
    primaryClass = "stat.CO",
    doi = "10.1007/s11222-018-9844-0",
    journal = "Stat. Comput.",
    volume = "29",
    number = "5",
    pages = "891--913",
    year = "2018"
}

@article{Belanger:2013oya,
    author = "Belanger, G. and Boudjema, F. and Pukhov, A. and Semenov, A.",
    title = "{micrOMEGAs$\_$3: A program for calculating dark matter observables}",
    eprint = "1305.0237",
    archivePrefix = "arXiv",
    primaryClass = "hep-ph",
    reportNumber = "LAPTH-023-13",
    doi = "10.1016/j.cpc.2013.10.016",
    journal = "Comput. Phys. Commun.",
    volume = "185",
    pages = "960--985",
    year = "2014"
}

@article{Boehm:2003hm,
    author = "Boehm, C. and Fayet, Pierre",
    title = "{Scalar dark matter candidates}",
    eprint = "hep-ph/0305261",
    archivePrefix = "arXiv",
    doi = "10.1016/j.nuclphysb.2004.01.015",
    journal = "Nucl. Phys. B",
    volume = "683",
    pages = "219--263",
    year = "2004"
}

@inproceedings{Caputo:2026pdw,
    author = "Caputo, Andrea and Essig, Rouven",
    title = "{The Dark Photon: a 2026 Perspective}",
    eprint = "2603.08430",
    archivePrefix = "arXiv",
    primaryClass = "hep-ph",
    month = "3",
    year = "2026"
}

@article{Rizzo:2018vlb,
    author = "Rizzo, Thomas G.",
    title = "{Kinetic Mixing and Portal Matter Phenomenology}",
    eprint = "1810.07531",
    archivePrefix = "arXiv",
    primaryClass = "hep-ph",
    reportNumber = "SLAC-PUB-17326",
    doi = "10.1103/PhysRevD.99.115024",
    journal = "Phys. Rev. D",
    volume = "99",
    number = "11",
    pages = "115024",
    year = "2019"
}

@article{Balan:2024cmq,
    author = "Balan, Sowmiya and others",
    title = "{Resonant or asymmetric: the status of sub-GeV dark matter}",
    eprint = "2405.17548",
    archivePrefix = "arXiv",
    primaryClass = "hep-ph",
    reportNumber = "TTP24-015, P3H-24-033",
    doi = "10.1088/1475-7516/2025/01/053",
    journal = "JCAP",
    volume = "01",
    pages = "053",
    year = "2025"
}

@article{Ibe:2019gpv,
    author = "Ibe, Masahiro and Kobayashi, Shin and Nakayama, Yuhei and Shirai, Satoshi",
    title = "{Cosmological constraint on dark photon from N$_{eff}$}",
    eprint = "1912.12152",
    archivePrefix = "arXiv",
    primaryClass = "hep-ph",
    reportNumber = "IPMU19-0188",
    doi = "10.1007/JHEP04(2020)009",
    journal = "JHEP",
    volume = "04",
    pages = "009",
    year = "2020"
}

@article{Sabti:2019mhn,
    author = "Sabti, Nashwan and Alvey, James and Escudero, Miguel and Fairbairn, Malcolm and Blas, Diego",
    title = "{Refined Bounds on MeV-scale Thermal Dark Sectors from BBN and the CMB}",
    eprint = "1910.01649",
    archivePrefix = "arXiv",
    primaryClass = "hep-ph",
    reportNumber = "KCL-2019-75",
    doi = "10.1088/1475-7516/2020/01/004",
    journal = "JCAP",
    volume = "01",
    pages = "004",
    year = "2020"
}

@article{Krnjaic:2025noj,
    author = "Krnjaic, Gordan",
    title = "{Testing Thermal-Relic Dark Matter with a Dark Photon Mediator}",
    eprint = "2505.04626",
    archivePrefix = "arXiv",
    primaryClass = "hep-ph",
    reportNumber = "FERMILAB-PUB-25-0234-T",
    month = "5",
    year = "2025"
}

@article{DAMIC-M:2025luv,
    author = "Aggarwal, K. and others",
    collaboration = "DAMIC-M",
    title = "{Probing Benchmark Models of Hidden-Sector Dark Matter with DAMIC-M}",
    eprint = "2503.14617",
    archivePrefix = "arXiv",
    primaryClass = "hep-ex",
    doi = "10.1103/2tcc-bqck",
    journal = "Phys. Rev. Lett.",
    volume = "135",
    number = "7",
    pages = "071002",
    year = "2025"
}

@article{DarkSide:2022knj,
    author = "Agnes, P. and others",
    collaboration = "DarkSide",
    title = "{Search for Dark Matter Particle Interactions with Electron Final States with DarkSide-50}",
    eprint = "2207.11968",
    archivePrefix = "arXiv",
    primaryClass = "hep-ex",
    reportNumber = "FERMILAB-PUB-22-566-ND-PPD-SCD",
    doi = "10.1103/PhysRevLett.130.101002",
    journal = "Phys. Rev. Lett.",
    volume = "130",
    number = "10",
    pages = "101002",
    year = "2023"
}

@article{PandaX:2025rrz,
    author = "Zhang, Minzhen and others",
    collaboration = "PandaX",
    title = "{Search for Light Dark Matter with 259 Days of Data in PandaX-4T}",
    eprint = "2507.11930",
    archivePrefix = "arXiv",
    primaryClass = "hep-ex",
    doi = "10.1103/rtnh-jn8s",
    journal = "Phys. Rev. Lett.",
    volume = "135",
    number = "21",
    pages = "211001",
    year = "2025",
    note = "[Erratum: Phys.Rev.Lett. 136, 069901 (2026)]"
}

@article{CRESST:2024cpr,
    author = "Angloher, G. and others",
    collaboration = "CRESST",
    title = "{First observation of single photons in a CRESST detector and new dark matter exclusion limits}",
    eprint = "2405.06527",
    archivePrefix = "arXiv",
    primaryClass = "astro-ph.CO",
    doi = "10.1103/PhysRevD.110.083038",
    journal = "Phys. Rev. D",
    volume = "110",
    number = "8",
    pages = "083038",
    year = "2024"
}

@article{Liu:2017htz,
    author = "Liu, Yu-Sheng and Miller, Gerald A.",
    title = {{Validity of the Weizs{\"a}cker-Williams approximation and the analysis of beam dump experiments: Production of an axion, a dark photon, or a new axial-vector boson}},
    eprint = "1705.01633",
    archivePrefix = "arXiv",
    primaryClass = "hep-ph",
    reportNumber = "NT@UW-17-05",
    doi = "10.1103/PhysRevD.96.016004",
    journal = "Phys. Rev. D",
    volume = "96",
    number = "1",
    pages = "016004",
    year = "2017"
}

@article{Alwall:2014hca,
    author = "Alwall, J. and Frederix, R. and Frixione, S. and Hirschi, V. and Maltoni, F. and Mattelaer, O. and Shao, H. -S. and Stelzer, T. and Torrielli, P. and Zaro, M.",
    title = "{The automated computation of tree-level and next-to-leading order differential cross sections, and their matching to parton shower simulations}",
    eprint = "1405.0301",
    archivePrefix = "arXiv",
    primaryClass = "hep-ph",
    reportNumber = "CERN-PH-TH-2014-064, CP3-14-18, LPN14-066, MCNET-14-09, ZU-TH-14-14",
    doi = "10.1007/JHEP07(2014)079",
    journal = "JHEP",
    volume = "07",
    pages = "079",
    year = "2014"
}

@article{LDMX:2023zbn,
    author = "{\r{A}}kesson, Torsten and others",
    collaboration = "LDMX",
    title = "{Photon-rejection power of the Light Dark Matter eXperiment in an 8 GeV beam}",
    eprint = "2308.15173",
    archivePrefix = "arXiv",
    primaryClass = "hep-ex",
    reportNumber = "FERMILAB-PUB-23-433-PPD-T, SLAC-PUB-17550",
    doi = "10.1007/JHEP12(2023)092",
    journal = "JHEP",
    volume = "12",
    pages = "092",
    year = "2023"
}

@article{Catena:2025fsl,
    author = "Catena, Riccardo and Gray, Taylor R. and Jerkvall, Thomas",
    title = "{Production of dark photons through higher electromagnetic moments at LDMX: simulations and model discrimination}",
    eprint = "2502.13635",
    archivePrefix = "arXiv",
    primaryClass = "hep-ph",
    doi = "10.1007/JHEP09(2025)040",
    journal = "JHEP",
    volume = "09",
    pages = "040",
    year = "2025"
}

@article{Rizzo:2023djp,
    author = "Rizzo, Thomas G.",
    title = "{Toward UV models of kinetic mixing and portal matter. V. Indirect probes of the new physics scale}",
    eprint = "2312.00226",
    archivePrefix = "arXiv",
    primaryClass = "hep-ph",
    reportNumber = "SLAC-PUB-17751",
    doi = "10.1103/PhysRevD.109.055039",
    journal = "Phys. Rev. D",
    volume = "109",
    number = "5",
    pages = "055039",
    year = "2024"
}

@article{Rizzo:2024bhn,
    author = "Rizzo, Thomas G.",
    title = "{Toward UV models of kinetic mixing and portal matter. VI. A more complex dark matter sector?}",
    eprint = "2408.01296",
    archivePrefix = "arXiv",
    primaryClass = "hep-ph",
    reportNumber = "SLAC-PUB-17779",
    doi = "10.1103/PhysRevD.110.075037",
    journal = "Phys. Rev. D",
    volume = "110",
    number = "7",
    pages = "075037",
    year = "2024"
}

@article{Planck:2018vyg,
    author = "Aghanim, N. and others",
    collaboration = "Planck",
    title = "{Planck 2018 results. VI. Cosmological parameters}",
    eprint = "1807.06209",
    archivePrefix = "arXiv",
    primaryClass = "astro-ph.CO",
    doi = "10.1051/0004-6361/201833910",
    journal = "Astron. Astrophys.",
    volume = "641",
    pages = "A6",
    year = "2020",
    note = "[Erratum: Astron.Astrophys. 652, C4 (2021)]"
}

@article{Gondolo:1990dk,
    author = "Gondolo, Paolo and Gelmini, Graciela",
    title = "{Cosmic abundances of stable particles: Improved analysis}",
    reportNumber = "UCLA-90-TEP-68",
    doi = "10.1016/0550-3213(91)90438-4",
    journal = "Nucl. Phys. B",
    volume = "360",
    pages = "145--179",
    year = "1991"
}

@article{DarkSHINE:2024guq,
    author = "Chen, Jing and others",
    collaboration = "DarkSHINE",
    title = "{DarkSHINE Baseline Design Report: Physics Prospects and Detector Technologies}",
    eprint = "2411.09345",
    archivePrefix = "arXiv",
    primaryClass = "physics.ins-det",
    month = "11",
    year = "2024"
}

@article{Speagle:2019ivv,
    author = "Speagle, Joshua S.",
    title = "{dynesty: a dynamic nested sampling package for estimating Bayesian posteriors and evidences}",
    eprint = "1904.02180",
    archivePrefix = "arXiv",
    primaryClass = "astro-ph.IM",
    doi = "10.1093/mnras/staa278",
    journal = "Mon. Not. Roy. Astron. Soc.",
    volume = "493",
    number = "3",
    pages = "3132--3158",
    year = "2020"
}

@article{Read:2002hq,
    author = "Read, Alexander L.",
    editor = "Whalley, M. R. and Lyons, L.",
    title = "{Presentation of search results: The $CL_s$ technique}",
    doi = "10.1088/0954-3899/28/10/313",
    journal = "J. Phys. G",
    volume = "28",
    pages = "2693--2704",
    year = "2002"
}

@article{Banerjee:2025ejz,
    author = "Banerjee, Avik and Catena, Riccardo and Gray, Taylor R.",
    title = "{Light Vector Dark Matter via a Magnetic Dipole Portal: Bridging Direct Detection and Fixed-Target Searches}",
    eprint = "2511.23259",
    archivePrefix = "arXiv",
    primaryClass = "hep-ph",
    month = "11",
    year = "2025"
}

@article{Catena:2023use,
    author = "Catena, Riccardo and Gray, Taylor R.",
    title = "{Spin-1 thermal targets for dark matter searches at beam dump and fixed target experiments}",
    eprint = "2307.02207",
    archivePrefix = "arXiv",
    primaryClass = "hep-ph",
    doi = "10.1088/1475-7516/2023/11/058",
    journal = "JCAP",
    volume = "11",
    pages = "058",
    year = "2023"
}

\end{document}